\documentclass[10pt,aps,prc,twocolumn,amsmath,floatfix,superscriptaddress,nofootinbib,preprintnumbers]
{revtex4-1}
\usepackage{color,graphicx}
\usepackage{mathptmx}                % math+times fonts
\usepackage{dcolumn}                 % Align table columns on decimal point
\usepackage{bm}                      % bold math
\usepackage{ulem,soul}
\usepackage{xcolor}
\usepackage{hyperref}

\allowdisplaybreaks

\begin{document}

\preprint{INT-PUB-17-030}

\title{The equation of state for dense nucleonic matter from a metamodeling. II. Predictions for neutron stars properties}

\author{J\'er\^ome Margueron}
\affiliation{Institute for Nuclear Theory, University of Washington, Seattle, Washington 98195, USA}
\affiliation{Institut de Physique Nucl\'eaire de Lyon, CNRS/IN2P3, Universit\'e de Lyon, Universit\'e Claude Bernard Lyon 1, F-69622 Villeurbanne Cedex, France}

\author{Rudiney Hoffmann Casali}
\affiliation{Institut de Physique Nucl\'eaire de Lyon, CNRS/IN2P3, Universit\'e de Lyon, Universit\'e Claude Bernard Lyon 1, F-69622 Villeurbanne Cedex, France}
\affiliation{Departamento de F\'isica, Instituto Tecnol\'ogico de Aeron\'autica, CTA, 12228900, S\~ao Jos\'e dos Campos, SP, Brazil}

\author{Francesca Gulminelli}
\affiliation{CNRS, ENSICAEN, UMR6534, LPC ,F-14050 Caen cedex, France}

\date{\today}

\begin{abstract}
Employing a recently proposed metamodeling for the nucleonic matter equation of state we analyze neutron star global properties such as masses, radii, momentum of inertia, and others. The impact of the uncertainty on empirical parameters on these global properties is analyzed in a  Bayesian statistical approach.
Physical constraints, such as causality and stability, are imposed on the equation of state and different hypotheses for the direct Urca (dUrca) process are investigated.
In addition, only metamodels with maximum masses above 2$M_\odot$ are selected.
Our main results are the following: the equation of state 
exhibits a universal behavior against the dUrca hypothesis under the condition of charge neutrality and $\beta$-equilibrium;
neutron stars, if composed exclusively of nucleons and leptons, have a radius of 12.7$\pm$0.4~km for masses ranging from 1 up to 2$M_\odot$; a small radius lower than 
11~km is very marginally compatible with our present knowledge of the nuclear empirical parameters;
and finally, the most important empirical parameters which are still affected by large uncertainties and play an important role in determining 
the radius of neutrons stars are the slope and curvature of the symmetry energy ($L_{sym}$ and $K_{sym}$) and, to a lower extent, the skewness parameters ($Q_{sat/sym}$).
\end{abstract}

\maketitle

%---------------------------------------------------------------------------------------
\section{Introduction}
\label{Sec: Introduction}

Neutron stars (NS) are the most compact stellar objects known to be lying on the stable branch
between white dwarfs and black holes~\cite{Book:Haensel2007}. 
Their radii are estimated to be $10$-$15$~km and their observed masses range between $1.2$ and $2M_\odot$.
As a consequence, their average density is about $10^{14\text{-}15}$~g~cm$^{-3}$ which is comparable
to the density of atomic nuclei.
The standard picture for NS composition therefore assumes that they are composed of neutrons and protons, embedded in 
a gas of electrons and muons.
NS matter is at $\beta$-equilibrium and, for a positive symmetry energy, this implies that neutrons are more
abundant than protons~\cite{Book:Haensel2007}.

More refined models for the NS interior composition assume various kind of phase transitions, from hyperonic matter
to quark matter.
The determination of the onset densities of these phase transitions requires an accurate
knowledge of the interaction among these particles, which is not attained yet.  
From the observational view point there is no clear signal indicating that NS inner cores contain exotic particles such as
hyperons or deconfined quarks.
In this work, we assume that matter is exclusively composed of neutrons, protons, electrons and muons and we will
predict the confidence intervals for various quantities related to NS properties, such as radii, masses, moment of inertia, etc...

We employ a metamodeling for the nuclear equation of state which we have introduced in Ref.~\cite{Margueron2017a}, hereafter called paper I.
The advantage of this approach for the NS equation of state  is that all possible predictions for dense and asymmetric nuclear
matter can be explored, provided they are compatible with nuclear physics knowledge of a few empirical parameters, 
like $E_{sat}$, $n_{sat}$, $K_{sat}$, $E_{sym}$, $L_{sym}$ or $K_{sym}$ (see paper~I for more details).
Another advantage of this metamodeling is that, {at variance with polytropic EOS,} matter composition is directly obtained from the
$\beta$-equilibrium, where the density dependence of the proton fraction could be obtained as function of the
model parameter.
In particular, different scenarii for the proton fraction in NS can be explored which impact the possibility for direct neutrino emission  (dUrca) fast cooling.
A link between the empirical parameters and  fast cooling will therefore be investigated in this work.
In addition, external constraints can be added to the metamodeling in order to filter out the parameterization exhibiting unphysical behavior. 
Examples of such constraints are the requirement that at least $2M_\odot$ can be reached by the NS meta-EOS, or that matter remains causal up to central densities of neutron stars with $2M_\odot$,
since the highest observed NS masses with small uncertainty are $1.667\pm0.021~M_\odot$ for PSR J1903+0327~\cite{Champion2008},
$1.928\pm0.017~M_\odot$ for PSR J1614-2230~\cite{0004-637X-832-2-167}
(initially measured to be $1.97\pm0.04~M_\odot$~\cite{Demorest2010a}) and 
$2.01\pm0.04~M_\odot$ for PSR J0348+0432~\cite{Antoniadis2013}.
For these reasons, the meta-EOS from Ref.~\cite{Margueron2017a}
offers a unique possibility to incorporate in the nuclear EOS the best knowledge issued from nuclear
physics, reducing the number of free parameters, and focusing on the most influential ones.
By varying these empirical parameters within reasonable ranges, accurate confidence intervals for the predictions of NS properties exclusively based on nucleonic matter can be obtained.

The present paper is organized as follows: in Sec.~\ref{Sec:empirical}, a short review of the meta-EOS from paper~I is performed and
the uncertainties on the empirical parameters are recalled.
Then the meta-EOS is implemented for $\beta$-equilibrium matter in NS and a simple perturbation analysis shows the impact of each
empirical parameter within its uncertainty on the mass-radius relation in Sec.~\ref{Sec:NSMR}.
In Sec.~\ref{sec:filters}, a more ambitious analysis is carried out based on Bayesian statistics, where  
a set of simple physical constraints are applied, such as causality, stability and positiveness of the symmetry energy.
In addition, we analyze consistently three hypothesis for the dUrca process, which directly depend on the density dependence of the symmetry energy.
The Bayesian analysis allows us to predict global properties of neutron stars as well as general density dependence of the EOS and its derivatives.
Finally, in Sec.~\ref{sec:inversion}, we address the inversion problem: how a measure of the mass and radius of a neutron star reflects in the
selection of the EOS, and which empirical parameters are mostly impacted?
Conclusions and outlooks are given in Sec.~\ref{Sec:Conclusions}.

%%%%%%%%%%%%%%%%%%%%%%%%%%%%%%%%%%%%%%%%%%%%%%%%%%%%%%%
\section{A metamodeling for the nuclear equation of state}
\label{Sec:empirical}
%%%%%%%%%%%%%%%%%%%%%%%%%%%%%%%%%%%%%%%%%%%%%%%%%%%%%%%

We briefly recall in this section the main features of the equation of state metamodeling which we use in this work.
We refer to paper~I for more details~\cite{Margueron2017a}.

Nuclear matter composed of neutrons and protons is characterized by the
isoscalar (is)  $n_0=n_n+n_p$ and isovector (iv)  $n_1=n_n-n_p$ densities, where $n_{n/p}$ is the neutron/proton density defined as a function of the Fermi momentum $k_{F_{n/p}}$ as,
\begin{equation}
n_{n/p} = %\frac 2 V \sum_{k<k_{F_{n/p}}} 1 = 
\frac{1}{3\pi^2} k_{F_{n/p}}^3 .
\end{equation}
Isospin asymmetric nuclear matter (ANM) can also be defined in terms of the asymmetry parameter
$\delta=n_1/n_0$, with the two boundaries $\delta=0$ and 1 corresponding to symmetric nuclear matter (SNM) and to 
pure neutron matter (PNM) respectively.
The saturation density of SNM is defined as the density at which the nucleonic pressure is zero and
it is denoted as $n_{sat}$.

The general properties of relativistic and non-relativistic nuclear interactions are often characterized in
terms of the nuclear empirical parameters, defined as the coefficients of the following series expansion in the
parameter $x=(n_0-n_{sat})/(3n_{sat})$~\cite{Piekarewicz2009},
\begin{eqnarray}
 e_{is} &=& E_{sat}+\frac{1}{2}K_{sat}x^{2}+\frac{1}{3!}Q_{sat}x^{3}+\frac{1}{4!}Z_{sat}x^{4}+...\, , 
 \label{eq:eis}\\
 e_{iv} &=& E_{sym}+L_{sym}x+\frac{1}{2}K_{sym}x^{2}+\frac{1}{3!}Q_{sym}x^{3} 
+\frac{1}{4!}Z_{sym}x^{4}+...\, , \nonumber \\
\label{eq:eiv}
 \end{eqnarray} 
where the isoscalar energy $e_{is}$ and the isovector energy $e_{iv}$ enter into the definition of
the energy per particle in nuclear matter, defined as
\begin{eqnarray}
 e(n_0,n_1)&=&e_{is}(n_0)+\delta^{2} e_{iv}(n_0).
 \label{eq:bindingenergy} 
\end{eqnarray} 
The isovector energy $e_{iv}$ is often called the symmetry energy $S(n_0)=e_{iv}(n_0)$.

In this work, we consider the metamodeling ELFc introduced in Ref.~\cite{Margueron2017a}.
In this metamodeling the energy per particle is defined as
\begin{eqnarray}
e^N(n_0,n_1)=t^{FG*}(n_0,n_1)+v^N(n_0,n_1).
\label{eq:ELFc}
\end{eqnarray}
where the kinetic energy reads,
\begin{eqnarray}
t^{FG^*}(n_0,n_1)&=&\frac{t_{sat}^{FG}}{2}\left(\frac{n_0}{n_{sat}}\right)^{2/3} 
\bigg[ \left( 1+\kappa_{sat}\frac{n_0}{n_{sat}} \right) f_1(\delta) \nonumber \\
&& \hspace{2.5cm} + \kappa_{sym}\frac{n_0}{n_{sat}}f_2(\delta)\bigg] ,
\label{eq:effmassms2}
\end{eqnarray}
and the potential energy is expressed as,
\begin{eqnarray}
v^N(n_0,n_1)=\sum_{\alpha\geq0}^N \frac{1}{\alpha!}( v_{\alpha}^{is}+ v_{\alpha}^{iv} \delta^2) x^\alpha u^N_{\alpha}(x).
\label{eq:vELFc2}
\end{eqnarray}
where $u^N_{\alpha}(x)=1-(-3x)^{N+1-\alpha}\exp(-bn_0/n_{sat})$ and $b=10\ln 2\approx 6.93$.
In Eq.~(\ref{eq:effmassms2}), the functions $f_1$ and $f_2$ are defined as
\begin{eqnarray}
f_1(\delta) &=& (1+\delta)^{5/3}+(1-\delta)^{5/3} , \\
f_2(\delta) &=& \delta \left( (1+\delta)^{5/3}-(1-\delta)^{5/3} \right) .
\end{eqnarray}

\begin{table*}[t]
\centering
\setlength{\tabcolsep}{2pt}
\renewcommand{\arraystretch}{1.2}
\begin{ruledtabular}
\begin{tabular}{ccccccccc}
$P_{\alpha}$ & $E_{sym}$ & $L_{sym}$ & $K_{sat}$ & $K_{sym}$ & $Q_{sat}$ & $Q_{sym}$ & $Z_{sat}$ & $Z_{sym}$ \\
                    & MeV           & MeV           & MeV           & MeV           & MeV           & MeV           & MeV           & MeV \\
\hline
$P_{\alpha,1}$ & 32 & 60 & 230 & -100 & 300 & 0 & -500 & -500 \\
$P_{\alpha,2}$ &  2 & 15 & 20 & 100 & 400 & 400 & 1000 & 1000\\
Min            & 26 & 20  &  190  & -400  &  -1300 &   -2000 &  -4500 & -5500 \\
Max             & 38 & 90  &  270  &  200  &   1900 &   2000 &   3500 &    4500 \\
step            &   2 & 10  &    20  &    75  &     400 &     400 &   1000 &    1000 \\
$N$            &   7 &   8  &      5  &      9  &       9 &       11 &       9 &        11 \\
\end{tabular}
\end{ruledtabular}
\caption{Characterization of the empirical parameters entering into the definition of the nuclear metamodeling ELFc. See text for more details.}
\label{tab:epbound}
%\end{center}
\end{table*}
%\end{sidewaystable}

The parameters  $\kappa_{sat/sym}$  can be directly expressed in terms of
the expected Landau effective mass at saturation density,
\begin{eqnarray}
\kappa_{sat}&=&\frac{m}{m_{sat}^*} - 1=\kappa_{s}, \; \hbox{ in SM ($\delta=0$)}  , \nonumber \\
\kappa_{sym}&=&\frac 1 2 \left[ \frac{m}{m^*_n} - \frac{m}{m^*_p} \right]=\kappa_{s}-\kappa_{v}, \;  \hbox{ in NM ($\delta=1$)} .
\end{eqnarray}
Fixing $\kappa_{sat/sym}$ to the expected values at saturation density,  
there is a one-to-one correspondence between the parameters $v_{\alpha}^{is}$ and
$v_{\alpha}^{iv}$ and the empirical parameters.  
We have for the isoscalar parameters,
\begin{eqnarray}
v_{\alpha=0}^{is} &=& E_{sat}-t_{sat}^{FG}(1+\kappa_{sat}), \label{eq:vis}\\ 
v_{\alpha=1}^{is} &=& -t_{sat}^{FG}(2+5\kappa_{sat}) ,\\ 
v_{\alpha=2}^{is} &=& K_{sat}-2t_{sat}^{FG}(-1+5\kappa_{sat}) , \\ 
v_{\alpha=3}^{is} &=& Q_{sat}-2t_{sat}^{FG}(4-5\kappa_{sat}) ,\\ 
v_{\alpha=4}^{is} &=& Z_{sat}-8t_{sat}^{FG}(-7+5\kappa_{sat}) ,
\\ \nonumber
\end{eqnarray}
and the isovector parameters,
\begin{eqnarray}
v_{\alpha=0}^{iv}&=&E_{sym}-\frac{5}{9}t_{sat}^{FG}[1+(\kappa_{sat}+3\kappa_{sym})] , \label{eq:viv}\\ 
v_{\alpha=1}^{iv}&=&L_{sym}-\frac{5}{9}t_{sat}^{FG}[2+5(\kappa_{sat}+3\kappa_{sym})] , \\ 
v_{\alpha=2}^{iv}&=&K_{sym}-\frac{10}{9}t_{sat}^{FG}[-1+5(\kappa_{sat}+3\kappa_{sym})] , \\ 
v_{\alpha=3}^{iv}&=&Q_{sym}-\frac{10}{9}t_{sat}^{FG}[4-5(\kappa_{sat}+3\kappa_{sym})] , \\ 
v_{\alpha=4}^{iv}&=&Z_{sym}-\frac{40}{9}t_{sat}^{FG}[-7+5(\kappa_{sat}+3\kappa_{sym})] . \label{eq:viv2}
\end{eqnarray}

Thanks to these relations, we can directly explore the impact of varying a single empirical parameter on the properties
on dense nucleonic matter and on the properties of neutron stars and supernovae matter. 
This allows making sensitivity analysis of the different parameters, and to avoid spurious correlations among them, which might be generated by a specific functional form. 
The price to pay for this flexibility is that 
{almost all correlations are suppressed, whether they are physical or unphysical.
In addition, since a lot of non-trivial density behavior is allowed, we have to take special care of each EOS.
This is done}
by applying filters on the model parameters based on different constraints from general physics and NS phenomenology.
{The final gain is that it becomes possible to control the link between the filters and the induced correlations, as we will see hereafter.}

In paper~I, we have analyzed the possible domain of variation for the empirical parameters.
The average values ($P_{\alpha,1}$) and they uncertainties ($P_{\alpha,2}$) are recalled in
Tab.~\ref{tab:epbound}.
$P_{\alpha,1}$ and $P_{\alpha,2}$ can be interpreted as the first (average) and second moment (standard deviation) of a Gaussian probability distribution for each of these parameters.
For this reason, $P_{\alpha,2}$ could be associated to the $1\sigma$ uncertainty, and the associated parameter may be
varied in a wider interval.
In lines 3 and 4 are written the max and min value associated to the largest exploration for the empirical parameters which is performed in the following.
The last lines 5 and 6 give the steps unit and the number of steps for our largest exploration.

Let us notice that some empirical parameters are not present in Tab.~\ref{tab:epbound}, such as $E_{sat}$, $n_{sat}$, 
$\kappa_{sat}$ and $\kappa_{sym}$.
The reason is that we have evaluated in paper~I that these empirical parameters are sufficiently well known and/or  have a very weak impact on the dense matter EoS.
For the simplicity of the discussion as well as to keep computing time within a reasonable range, we have decided to fix the value of these
parameters in this work to be: $E_{sat}=-15.8$~MeV, $n_{sat}=0.155$~fm$^{-3}$, $\kappa_{s}=0.3333$ and $\kappa_{v}=0.4218$.
This choice leads to the Landau effective mass in symmetric matter $m^*_{sat}/m=0.75$ and its isospin splitting $\Delta m^*_{sat}=0.1$,
see paper~I for more details.

%%%%%%%%%%%%%%%%%%%%%%%%%%%%%%%%%%%%%%%%%%%%%%%%%%%%%
\section{Neutron star masses and radii}
\label{Sec:NSMR}
%%%%%%%%%%%%%%%%%%%%%%%%%%%%%%%%%%%%%%%%%%%%%%%%%%%%%

In this section, we investigate how sensitive are the NS masses and radii measurements on the nuclear empirical
parameters defined by Eqs.~(\ref{eq:eis})-(\ref{eq:eiv}).
To do so, we solve the hydrostatic equations in general relativity for spherical and non-rotating stars, also
named the TOV equations~\cite{Tolman1939,Oppenheimer1939,Book:Haensel2007},
\begin{eqnarray}
\frac{d m(r)}{d r} &=& 4\pi r^2 \rho(r) , \nonumber \\
\frac{d P(r)}{d r} &=& -\rho c^2 \left( 1+\frac{P}{\rho c^2}\right)\frac{d \Phi(r)}{d r} \, , \nonumber \\
\frac{d \Phi(r)}{d r} &=& \frac{G m}{c^2 r^2} \left(1+\frac{4\pi P r^3}{mc^2}\right)
\left(1-\frac{2Gm}{rc^2}\right)^{-1}\, , 
\label{eq:tov}
\end{eqnarray}
where $G$ is the gravitational constant, $P$ the pressure, $m(r)$ is the enclosed mass at the radius $r$, defined within the Schwarzschild 
metric $ds^2=e^{2\Phi}dct^2-e^{2\lambda}dr^2-r^2(d\theta^2+\sin^2\theta d\phi^2)$.
Both $\Phi$ and $\lambda$ are functions of $r$.
$\Phi$ is the gravitational potential and $e^{-2\lambda}=1-Gm/(rc^2)$.
Let us remark that $\rho$ in Eqs.~(\ref{eq:tov}) is the energy density containing a contribution from the rest mass %energy ($n_0$) 
and from the energy per particle $e$ as, $\rho c^2=(m_Nc^2 + e) n_0$.
Numerically, $m(r)$ and $P(r)$ are solved from 0 to $R$, fixing the boundary condition $m(0)=0$ and $P(0)=P_c$ where $P_c(\rho=\rho_c)$ 
is arbitrarily fixed. The variation of the central density $\rho_c$ generates a family of solutions with different $M$ and $R$, where $M=m(R)$, and the radius $R$ is defined 
as the radial coordinate for which $P(r=R)=0$.
Then $\Phi(r)$ is integrated from $R$ down to $r=0$, matching with the external solution $\Phi(r\ge R)=-\lambda(r\ge R)$.

\begin{figure*}[tb]
\begin{center}
\includegraphics[angle=0,width=0.7\linewidth]{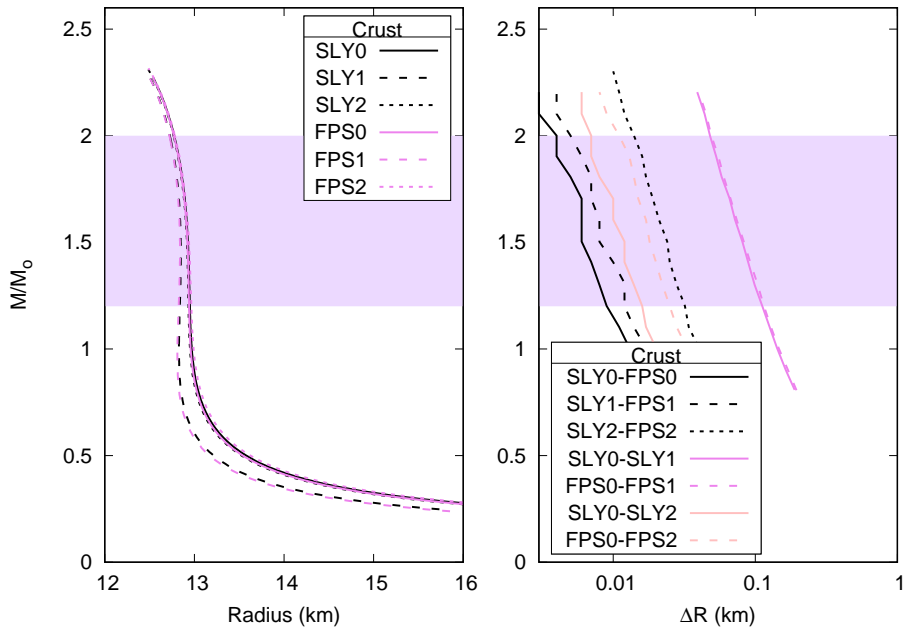}
\end{center}
\caption{(Color online) Effect of the crust and of the matching between the crust and the core EoS.
Left part: mass-radius relationship with different prescriptions for the choice of the crust EOS and of the matching procedure (see text). 
Right part: sensitivity analysis of the shift in the NS radius induced by the different choices given in the legend of the figure (see text also).
For instance, the solid black line stands for $\Delta R=R($SLY0$)-R($FPS0$)$.
The largest impact is found for the modification of the low-density boundary of the matching between the crust and the core.
Even for this extreme case, the uncertainty in the radius definition is less than 0.1~km.}
\label{fig:MRcrust} 
\end{figure*}

For a non-rotating NS the surface gravitational redshift $z$ is simply defined as~\cite{Haensel1982}
\begin{equation}
z = \left(1-\frac{2GM}{Rc^2}\right)^{-1/2} - 1 =  \left(1-\frac{R_s}{R}\frac{M}{M_{\odot}}\right)^{-1/2} - 1 \, ,
\end{equation}
where $R_s$ is the Schwarzschild radius, $R_s=2.955$~km for $M=M_\odot$.

Considering slow and rigid rotation of the neutron star, the moment of inertia can be estimated from the lowest-order perturbative 
approximation~\cite{Hartle1967a,Morrison2004,Book:Haensel2007}.
The slow rotation approximation implies that centrifugal forces are small compared to the gravity, $R^3\Omega^2/(GM)\ll 1$,
where the angular frequency $\Omega$ is measured by a distant observer.
Notice that for the fastest observed pulsar PSR J1748-2446ad at 716 Hz (spin period of 1.396~ms)~\cite{Hessels2006},
we get $R^3\Omega^2/(GM)\approx 0.11$ assuming $M=1.4M_{\odot}$ and $R=10$~km.

The GR moment of inertia $I$ is given by the following expression~\cite{Hartle1967a,Morrison2004}:
\begin{eqnarray}
I=\frac{8\pi}{3}\int_0^R dr r^4 \rho \left(1 + \frac{P}{\rho c^2} \right)\frac{\bar{\omega}}{\Omega} e^{\lambda-\Phi} \, ,
\label{eq:I}
\end{eqnarray}
where $\bar{\omega}$ is the local spin frequency, which represent the GR correction to the asymptotic angular momentum $\Omega$. 
$\bar{\omega}$ is usually a small correction for NS and the local angular momentum reads $\omega=\Omega-\bar{\omega}$.
The familiar Newtonian expression for the moment of inertia can be recovered imposing $\lambda=\Phi=0$ and $P\ll \rho c^2$.
In practice, we first solve the static TOV equations~(\ref{eq:tov}), and then obtain the moment of inertia from Eq.~({\ref{eq:I}) fixing
$\bar{\omega}$ to be arbitrarily small at the center of the star.

\begin{figure*}[tb]
\begin{center}
\includegraphics[angle=0,width=0.7\linewidth]{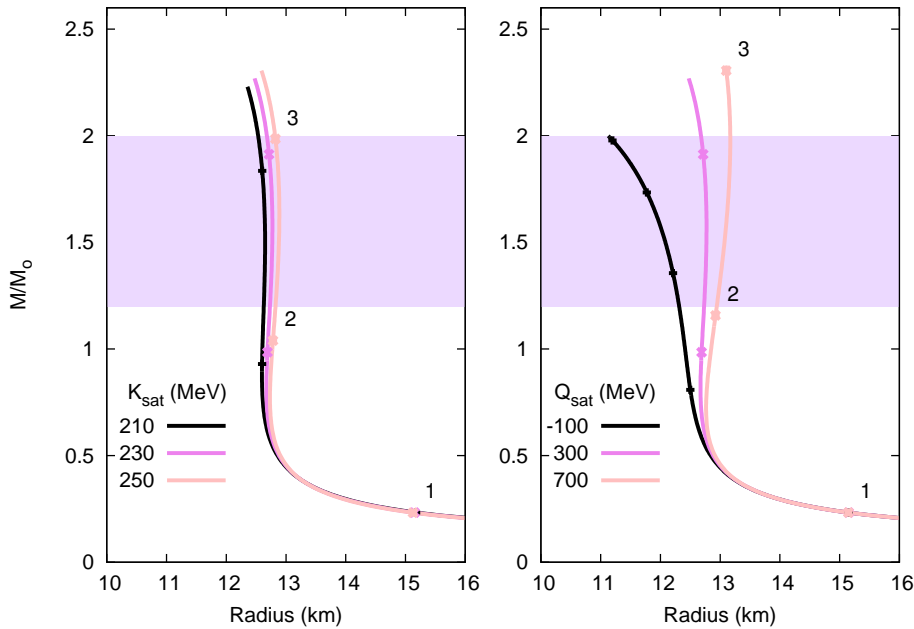}
\end{center}
\caption{(Color online) M-R diagram generated with meta-EOS where the isoscalar empirical parameters are varied.
See text for more details.}
\label{fig:MRsat} 
\end{figure*}

\subsection{The matching of the core and crust EOS}

In the core of NS, the dense matter EOS is composed of neutrons, protons, electrons and muons at $\beta$-equilibrium in the mean field
generated by the meta-EOS, see paper~I~\cite{Margueron2017a}.
Below saturation density, the core meta-EOS is matched to the EOS for the crust based on a cubic spline.
There is a discussion concerning the impact of the matching procedure on the NS radius~\cite{Fortin2016}.
The optimal matching should be performed with EOSs which have been consistently derived in the core and in the crust.
The derivation of the crust EOS by extending  our meta-functional to finite nuclei is currently in progress~\cite{Chatterjee2017}. 
For the present study, we have chosen to perform a log$\rho$-log$P$ cubic spline which guides the continuous
interpolation between the crust and the core EOS in the transition region.
To so do, we have to reserve a rather large region where the spline can smoothly connect the crust and the core.
We therefore stop the crust EOS at a density $n_0^{l}=0.1n_{sat}$ and start the core meta-EOS at the density $n_0^h=n_{sat}$.

We have estimated the accuracy of our prescription against the change of the crust EOS as well as against the values for
$n_0^l$ and $n_0^h$. 
For the crust EOS, we have considered two existing and widely used EOS, hereafter called SLY and FPS.
SLY is based on the Skyrme nuclear interaction SLy4~\cite{Chabanat1998a} which has been applied for the crust EOS
considering a compressible-liquid-drop-model~\cite{Douchin2001a}.
FPS is the crust EOS from Ref.~\cite{Pandharipande1989}.
The properties derived from these EOS consistently matched in the core are discussed in Ref.~\cite{Haensel2004}.
We use tables provided by the IOFFE institute and available on-line~\footnote{http://www.ioffe.ru/astro/NSG/NSEOS/}.
We now discuss the results which are shown in Fig.~\ref{fig:MRcrust}.

In Fig.~\ref{fig:MRcrust}, we discuss the effect of changing the crust EoS (considering SLY and FPS) and 
changing the density parameters $n_0^l$ and $n_0^h$.
The core meta-EoS is determined from the average empirical parameters given in line 1 of Tab.~\ref{tab:epbound}.
The pink-colored region in Fig.~\ref{fig:MRcrust} stands for the observed masses, e.g. between
1.2~M$_{\odot}$ and 2.0~M$_{\odot}$.
Modelings with number 0 represent the reference calculation: the curves SLY0 and FPS0 are obtained for the standard 
choice for $n_0^l$ and $n_0^h$: $n_0^{l}=0.1n_{sat}$ and $n_0^h=n_{sat}$.
Then we have increased the low density boundary, $n_0^{l}=0.2n_{sat}$, for SLY1 and FPS1, or decreased the high density
boundary, $n_0^h=0.7n_{sat}$, for SLY2 and FPS2.
Since the validity of the approach is based on the possibility to perform a cubic interpolation between $n_0^l$ and $n_0^h$
in the log$\rho$-log$P$ space, these boundary must be kept well separated.
We see from Fig.~\ref{fig:MRcrust} that changing the crust EoS has an impact on the predicted radius which is less than 20m, similar to changing the
value of $n_0^h$, and that the largest impact is for changing $n_0^l$ from $0.1n_{sat}$ to $0.2n_{sat}$.
For the latter case, the impact is estimated to be about 100m for low mass NS and about 50m for high mass NS.
We conclude that the uncertainties induced by the crust EoS in terms of which model to use and what are the matching
densities $n_0^l$ and $n_0^h$, induce an uncertainty in the crust thickness which is about 100m for low mass NS and 50m for high mass NS.

For the present discussion,  we consider that the present matching procedure based on log$\rho$-log$P$ cubic spline is sufficiently accurate for quantitative predictions on NS radii, considering their large observational uncertainty.
In the following, we set the crust EOS to be SLY and the reference core meta-EOS to be given by the average parameters given in Tab.~\ref{tab:epbound}.

\begin{figure*}[tb]
\begin{center}
\includegraphics[angle=00,width=0.9\linewidth]{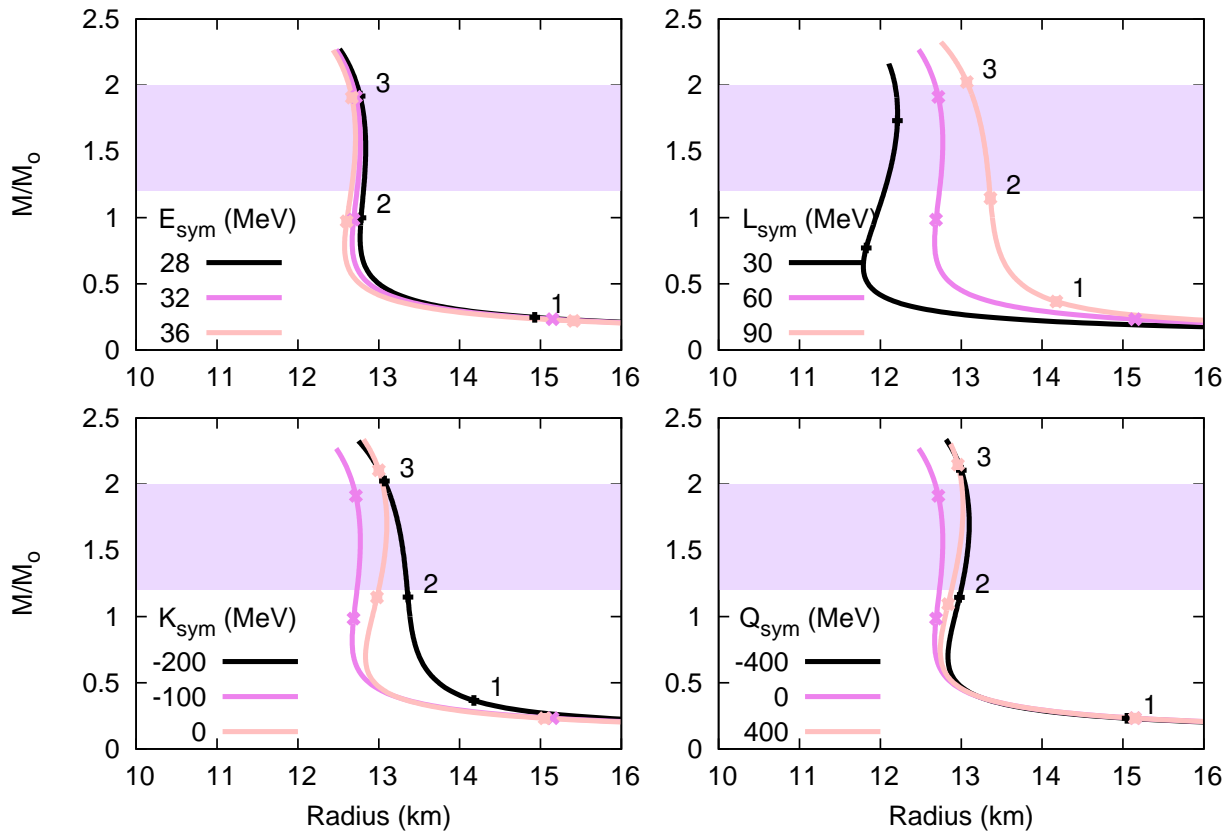}
\end{center}
\caption{(Color online) Same as Fig.~\ref{fig:MRsat} for the variation of isovector empirical parameters.}
\label{fig:MRsym} 
\end{figure*}

\subsection{Sensitivity analysis on the isoscalar and isovector empirical parameters}

The expected uncertainties for the empirical parameters are given in the second line of Tab.~\ref{tab:epbound}.
They have been obtained from the analysis presented in paper~I.
In this section, we discuss the impact of varying the isoscalar and isovector empirical parameters within these ranges. 
The advantage of the meta-EOS is that we can directly measure the impact of changing only one of the empirical parameter inside its range of uncertainty, without changing the other parameters. 
Here we analyze the impact of these uncertainties on the mass-radius relationship.

We represent in Fig.~\ref{fig:MRsat}  the impact of the isoscalar empirical parameters
$K_{sat}$ (incompressibility modulus) and $Q_{sat}$ (skewness), the impact of $E_{sat}$ and 
$n_{sat}$ being extremely weak.
The crosses represent the value of the central density in units of $n_{sat}$.
It can  be observed that as the value of the empirical parameter $K_{sat}$ or $Q_{sat}$ increases, 
the EOS becomes stiffer, and the NS radius consequently increases.
The impact of varying $K_{sat}$ on the radius is quite weak: it goes from about 200~m at the
low mass boundary up to 400~m for the high mass, and is about 300~m at the canonical mass.
The skewness parameter $Q_{sat}$ is found to have a larger impact, mostly because the
value of this parameter is yet rather unconstrained.
The low value of $Q_{sat}$ considered here (-100~MeV) is found to produce a very soft EOS which
can reach 2$M_\odot$ for a central density of about $5n_{sat}$.
Beyond 2$M_\odot$ this EOS is too soft and predicts $(v_s/c)^2<0$, where $v_s$ is the sound velocity.

We now discuss the impact of the isovector empirical parameters $E_{sym}$ (symmetry energy), $L_{sym}$
(slope), $K_{sym}$ (curvature) and $Q_{sym}$ (skewness) on the mass-radius relationship in Fig.~\ref{fig:MRsym}.
The impact of $E_{sym}$ is small, as expected, since it does not impact the pressure, and has a weak effect on the energy density.
The uncertainty on this parameter is also rather small compared to the others (note that we have considered a 2$\sigma$
variation for this parameter).
The impact of $L_{sym}$ and $K_{sym}$ is clearly larger.
We recall that $L_{sym}$ and $K_{sym}$ were identified as being the main source of uncertainty for the nuclear
EOS in paper~I~\cite{Margueron2017a}.
The uncertainty of $L_{sym}$ leads to an uncertainty of about 2~km at low NS mass and 1~km at high NS mass,
about 1.5~km at canonical mass.
The effect of $K_{sym}$ is also quite large: about 1~km at low NS mass and 500~m at high NS mass.
it is interesting to note that the largest impact of the uncertainty of $L_{sym}$ is found to be for masses below 1.2$M_\odot$
where the densities are between $n_{sat}$ and $2n_{sat}$, while the impact of $K_{sym}$ is found to be biggest
at slightly larger masses.
This is a consequence of the Taylor expansion which is at the base of our theoretical modeling: the impact of the
higher order empirical parameters is larger at higher density, while the lower order empirical parameters are more important 
around saturation density $n_{sat}$.
Since $L_{sym}$ and $K_{sym}$ are the main source of uncertainties in the nucleon pressure, see paper~I~\cite{Margueron2017a} for instance, our present analysis is compatible with the empirical $RP^{-1/4}$ correlation from Ref.~\cite{Lattimer2001a}, see also Fig.~\ref{fig:impactPR} and the discussion
at the end of Sec.~\ref{sec:impactns}.

The impact of $Q_{sym}$ remains small and of the order of the uncertainty on the incompressibility modulus $K_{sat}$.
Despite the very large uncertainties on this parameter, the reason of its weak impact is that the densities at which this 
empirical parameter plays a role are still above the largest densities considered here.

In conclusion, we observe that the largest impact on the mass-radius relationship is
given by the three empirical parameters $Q_{sat}$, $L_{sym}$ and $K_{sym}$.
Better estimations of these parameters may come from nuclear physics experiments such as
as PREX and CREX~\cite{Abrahamyan2012}, as well as very precise measurement of collective modes in nuclei
such as the GDR, GQR, see Ref.~\cite{Colo2014} for instance. Detailed discussion on nuclear experimental investigations can
be found in Refs.~\cite{BALi2013a,Horowitz2014,BALi2014}.
They may also come from better knowledge of the NS radii, which are nowadays intensively investigated from various approaches:
X-ray emission from quiescent low-mass X-ray binaries (qLMXB)~\cite{Guillot2011,Guillot2013,Guillot2014,Heinke2014,Guillot2016,Bogdanov2016},
from observation of photospheric expansions in X-ray burst (XRB)~\cite{Ozel2009,Ozel2010,Nattila2016},
or from the precise X-ray timing of millisecond pulsars enabled by the upcoming observations with the NICER mission~\cite{Gendreau2012}.

%%%%%%%%%%%%%%%%%%%%%%%%%%%%%%%%%%%%%%%%%%%%%%%%%%%%%
\section{Confronting to physical constraints}
\label{sec:filters}
%%%%%%%%%%%%%%%%%%%%%%%%%%%%%%%%%%%%%%%%%%%%%%%%%%%%%

The sensitivity analysis presented in the previous section gives a clear understanding of the parameters which should be better constrained to improve our understanding of NS, but it cannot be considered as a quantitative estimation of error bars on the astrophysical quantities. 
Indeed, the different empirical parameters can be correlated, meaning that to estimate the impact of their uncertainty they have to be varied collectively in the full parameter space.
Moreover, {as noticed in paper~I}, the parameter space is so large that 
among all the considered EOS, some may violate some basic requirements, such as causality for instance. 
The domain of variation of the empirical parameters may therefore be reduced by imposing some basic physical requirements.
In this section, we apply several filters to the explored meta-EOS, namely:
\begin{itemize}
\item stability: the gradient of the pressure and of the energy density $\rho$ (with mass terms) should be  positive at all densities ;
\item symmetry energy: the symmetry energy $S(n_0)$ should be positive at all densities.
\item causality: the speed of sound $v_s$ should not exceed the speed of light ($v_s^2<c^2$) and we exclude imaginary values ($v_s^2<0$) as well ;
\end{itemize}
These requirements are imposed along the $\beta$-equilibrium path and for a density interval defined between $n_{sat}$ and $n_{max}$ 
where $n_{max}$ is the central density for a NS with $M=2M_{\odot}$. 
The density $n_{max}$ is calculated for each meta-EOS and all the meta-EOS which do not reach $2M_{\odot}$ are also rejected.
The sound velocity is calculated considering n, p, e$^-$, and $\mu^-$ as,
\begin{eqnarray}
\left(v_{s}/c\right)^2 = \frac{d P_{tot}}{d n_0}/\left[ m_{tot}c^2+e_{tot}+P_{tot}/n_0 \right]\, ,
\label{eq:vs}
\end{eqnarray}
where $m_{tot}=x_pm_pc^2+x_nm_nc^2+x_em_ec^2+x_\mu m_\mu c^2$. $e_{tot}$ and $P_{tot}$ are the total
energy per particle and pressure, and $x_n$, $x_p$, $x_e$ and $x_\mu$ are particle fractions for n, p, e$^-$, and $\mu^-$.

Let us discuss briefly the expected behavior of the symmetry energy above saturation density.
There is a tight link between the symmetry energy above saturation density and fast cooling induced by the dUrca process.
The dUrca process (direct neutrino emission) is based on the fact that neutron star in $\beta$-equilibrium balances the following reactions:
\begin{eqnarray}
n \rightarrow p+e^{-}+{\bar \nu}_e \, , \;\;
p+e^{-} \rightarrow n+{\nu}_e \, .
\end{eqnarray}
The second process (electron capture) is {Pauli blocked, except if }
protons and electrons are sufficiently energetic.
This implies that the proton Fermi momentum, and therefore the proton density, must be sufficiently high for the electron capture to occur. 
If not, $\beta$-equilibrium is insured by more complex weak interactions that involve a higher number of particles (modified Urca), and therefore happen with a much slower rate~\cite{Lattimer1991}.
The dUrca process is  possible if the proton fraction $x_p>1/9$ in n, p, e$^{-}$ matter~\cite{Lattimer1991}.
In the presence of muons, the dUrca condition is slightly modified to be~\cite{Klahn2006}:
\begin{eqnarray}
x_p>x_{DU}\hbox{, where } \;
x_{DU} = \left[ 1 + (1+x_{ep}^{1/3})^3 \right]^{-1},\label{eq:durca}
\end{eqnarray}
and $x_{ep}=n_e/n_p=n_e/(n_e+n_\mu)$. 
In the absence of muons, we have $x_{ep}=1$ and the limit $x_{DU}=1/9$ is recovered~\cite{Lattimer1991}.
There is therefore a straightforward relation between enhanced cooling induced by dUrca process and the symmetry energy
$S(n_0)$ which governs the density dependence of the proton fraction. 
The neutrino emissivity deduced from the integration of the cross section over the phase space for dUrca gives a characteristic $T^6$ temperature dependence, 
{while it is suppressed as $T^8$ for mUrca,} see Ref.~\cite{Yakovlev2004a} and references therein.

This discussion is somewhat schematic because other complex mechanisms enter in the thermal properties of NS: in particular strong superfluidity and superconductivity in the core in pairing channels which are still poorly known~\cite{Leinson2015}, 
local magnetic fields in the crust~\cite{Bonanno2014}, 
Fermi surface depletion due to short-range nuclear correlations~\cite{Dong2016}, 
are all phenomena that might have an influence on the fast cooling scenario. 
Still, the relative weight of these mechanisms is not yet completely clear.
{Ultimately, a more complete analysis shall put all these effects together.
% to be able conclude more firmly.
At present, it is however still too ambitious, and}
we will keep Eq.~(\ref{eq:durca}) as the unique condition for the fast cooling to happen, see Refs.~\cite{Steiner2005a,BALi2008,Horowitz2014,BALi2014} for recent reviews.

{From the observational view point, there} are different classes of NS where enhanced cooling {might be critical:}
the cooling of isolated NS~\cite{Yakovlev2004a}, the cooling of magnetars~\cite{Vigano2013} and the thermal relaxation of Transient low-mass x-ray binaries (LMXBs)~\cite{Heinke2007}.
Yet isolated NS are compatible with the so-called minimal cooling scenario in which the dUrca process is excluded~\cite{Page2004,Page2009}, 
but it should be noticed that enhanced cooling would make most of isolated NS so cold that they would not be observed.
Indeed, less than half of the supernovae remnants within 5 kpc have identified central sources~\cite{Kaplan2004, Kaplan2006}.
The thermal luminosity of magnetars is systematically higher than that of classical pulsars, showing again the important role of the magnetic field
in the cooling process. Recent simulations of the cooling of magnetars have shown that the effect of the magnetic field is able to screen an
eventual fast cooling, keeping the temperature of magnetar rather high even if fast cooling is possible~\cite{Vigano2013}.
So firm conclusions concerning the impossibility of enhanced cooling could not be drawn yet {from isolated NS nor magnetars}.
Concerning qLMXBs, most of them are consistent with having standard cooling, however, with two exceptions: SAX J1808.4-3658~\cite{Heinke2007} 
and 1H 1905+00~\cite{Jonker2007}.
These two exceptions are extremely cold neutron stars for which only upper limit of the thermal component of the luminosity are reported.
In these cases, very low core temperatures may be explained by fast cooling.
Finally, NS do not only vary by their magnetic field, but also by their mass. 
For low-magnetic NS, the existence or not of dUrca process could be explained by their different masses. 
If the NS mass is too low, its central density is not high enough to reach the proton fraction threshold $x_{DU}$~\cite{Yakovlev2004a}.
In this scenario only high mass NS could be fast cooled; but to date, the masses of these NS are yet unknown. 
It is therefore difficult to estimate the critical mass above which dUrca is switched on.

In summary, the situation is the following: the dUrca process is certainly not possible for most of the NS, but some of them may be cooling rapidly.
The parameter which controls the cooling of NS may be the total mass, but we do not know what is the threshold mass allowing dUrca.
{Given these uncertainties, we decide to explore different scenarii for the dUrca condition threshold and investigate consistently their implications.}
The first one (DURCA-0) assumes that within the range of observed NS ($M<2M_\odot$), dUrca is not possible.
The second one (DURCA-1) assumes that only a few of the more massive observed NS could experience dUrca: {the dUrca threshold is crossed at least once in the range $1.8<M/M_\odot<2$.}
And finally, the last scenario (DURCA-2) assumes 
a lower range $1.6<M/M_\odot<1.8$ for the dUrca threshold to be satisfied.
These scenarii are mutually excluding each others.
%The predictions above $2M_{\odot}$ are disregarded since it is yet unknown if $2M_{\odot}$ neutron stars exist.
Neutron stars with more than $2M_{\odot}$ are disregarded.
In the following, we will compare the prediction based on these three hypothesis and check whether the static structure of NS is influenced by
these assumptions, which will be interesting since it will offer the possibility to check it against observations. 
If no modification is seen, this  will reveal a kind of universal behavior independently of the proton fraction inside the NS.

Let us now detail the selection of the models.
Since we already checked the very weak influence of some empirical parameters on the equation of state, such as for instance $n_{sat}$ and $E_{sat}$, as well as
the parameter governing the density and isospin asymmetry dependence of the effective mass ($\kappa_{sat}$ and $\kappa_{sym}$), see paper ~I~\cite{Margueron2017a}, 
we decide not to vary them in the present analysis.
%They are fixed to the following values, see Tab.~\ref{tab:epbound}: $n_{sat}=0.155$~fm$^{-3}$ and $E_{sat}=15.8$~MeV, $m^*/m=0.75$ and $\Delta m^*=0.1$. 

In the following, the 8 parameters ($E_{sym}$, $L_{sym}$, $K_{sat}$, $K_{sym}$, $Q_{sat}$, $Q_{sym}$, $Z_{sat}$ and $Z_{sym}$) are varied uniformly between their
minimum and maximum values given in Tab.~\ref{tab:epbound}.
Each parameter-set defines a different equation of state which properties are analyzed. 
For each parameter-set which satisfies all the physical requirements listed above, a probability $w_\mathrm{filter}=1$ is attributed, while if one or more physical 
requirements are violated, we set $w_\mathrm{filter}=0$. 
Taking advantage of the Bayesian approach, the probability $w_\mathrm{filter}$ is associated to the likelihood probability $p_{lik}$ as
\begin{eqnarray}
p_{lik} (\{P_{\alpha}\}_i) = \frac{1}{N_{lik}} \, w_\mathrm{filter} (\{P_{\alpha}\}_i) \, \prod_{\alpha=1}^8 \; g_{P_{\alpha,1},P_{\alpha,2}}(P_\alpha) \, ,
\label{eq:probalikely}
\end{eqnarray}
where the functions $g$ are the prior probabilities given by a Gaussian distribution,
\begin{eqnarray}
g_{P_{\alpha,1},P_{\alpha,2}}(P_\alpha) = \frac{1}{\sqrt{2\pi}P_{\alpha,2}} \exp{\Big[-\frac1 2 \left( \frac{P_\alpha-P_{\alpha,1}}{P_{\alpha,2}} \right)^2 } \Big]\, ,
\label{eq:probaprior}
\end{eqnarray}
and $P_{\alpha,1}$ and $P_{\alpha,2}$ are the average and standard deviation
of the prior-distribution of the $P_\alpha$ parameters, which are given in the two first lines of Tab.~\ref{tab:epbound}.
The normalization $N_{lik}$ is calculated by integrating the probability $p_{lik}$ over all the parameters $P_\alpha$: $N_{lik}=\int dP_1\cdots \int dP_8 \, p_{lik}$. 

The probability distribution $p_{lik}$ is focused where it is expected to be the most relevant. 
For the parameters where the uncertainty could be related to nuclear experimental knowledge, the prior distribution is an effective way to include this knowledge
in the present analysis. This is the case for instance for the parameters $E_{sym}$, $L_{sym}$ and $K_{sat}$.
The other parameters $K_{sym}$, $Q_{sat/sym}$ and $Z_{sat/sym}$ are not yet very well constrained by nuclear physics. 
Their uncertainties have instead been estimated from the predictions of various modelings, see paper~I~\cite{Margueron2017a}. 
For this reason, the uncertainties attributed to these parameters are arbitrary.
In consequence, we have considered a rather large domain of variation for these parameters covering 4$\sigma$ around the central value.
The details of the parameter mesh are given in Tab.~\ref{tab:epbound}: minimum and maximum value as well as the step unit and number of steps considered. 
In total it is about 25 millions of EOS which are analyzed.
This massive computational work was made possible using the CC-IN2P3 super-computing facility which dedicated about 500 CPU for one-month.

From the 25 millions of initial EOS, there are finally 4 millions satisfying the physical requirements as well as the DURCA-0 hypothesis, about 600,000 for DURCA-1 and 700,000 for DURCA-2.

\begin{figure*}[tb]
\begin{center}
\includegraphics[angle=0,width=1.0\linewidth]{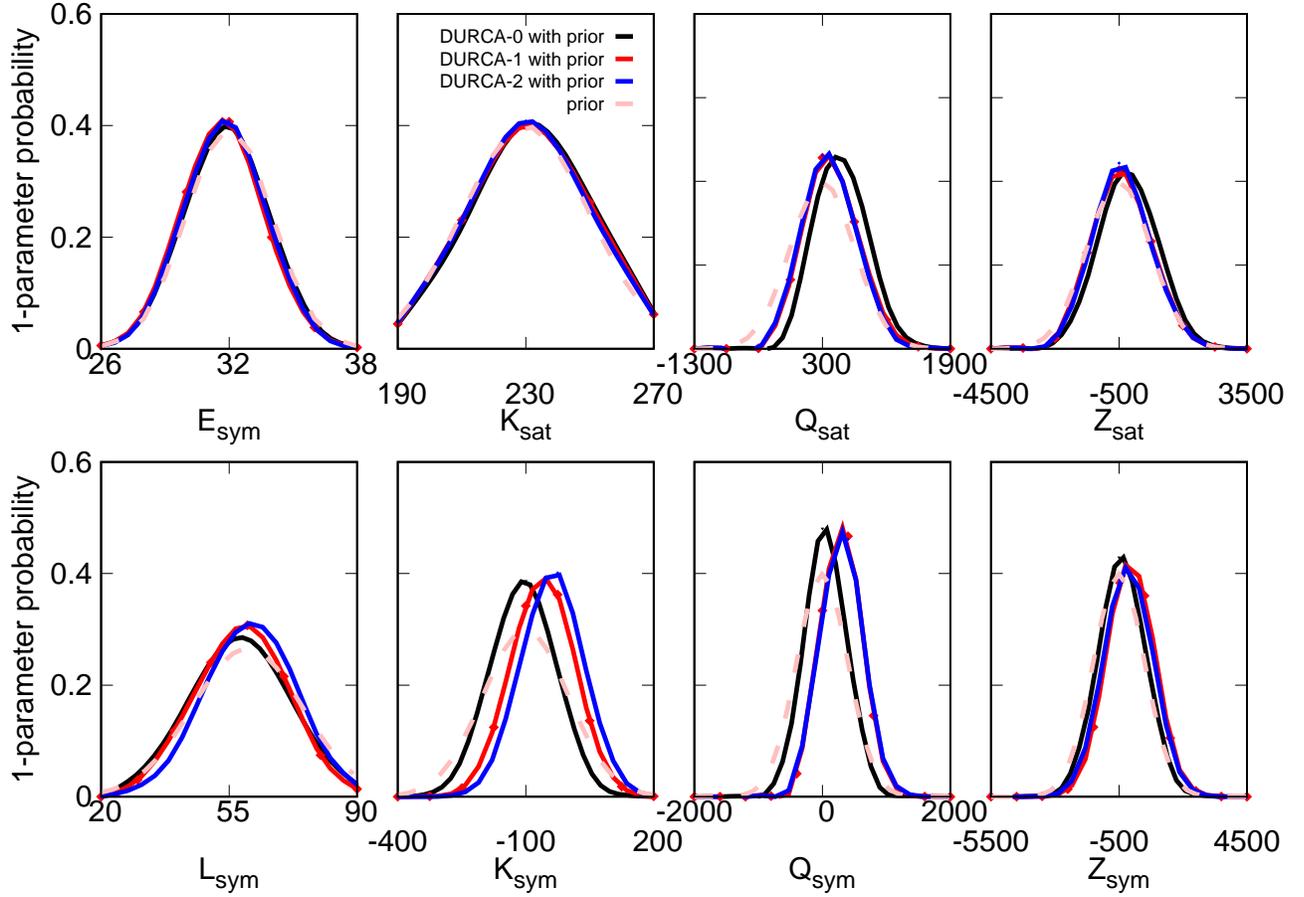}
\end{center}
\caption{(Color online) 1-parameter probability distributions $p(P_\alpha)$. See text for more details.}
\label{fig:prob} 
\end{figure*}

\subsection{Analysis of $p_{lik}$}
\label{sec:impactep}

Since it is impossible to visualize the probability $p_{lik}$ in the 8 dimensional  parameter space, the reduced 1 and 2-parameters probabilities are introduced as:
\begin{eqnarray}
p_1(P_\alpha) &=& \left\{\prod_{\beta(\ne\alpha)=1}^8 \int dP_\beta\right\} \; p_{lik} (\{P_{\beta}\}) \, ,  \\
p_2(P_\alpha,P_\beta) &=&  \left\{\prod_{\gamma(\ne\alpha,\beta)=1}^8 \int dP_\gamma\right\} \; p_{lik} (\{P_{\gamma}\}) \, .
\end{eqnarray}
The 1-parameter probability $p_1(P_\alpha)$ allows to visualize where the domain of solutions for each empirical parameter is located after filtering, while the 2-parameters probability $p_2(P_\alpha,P_\beta)$ allows recognizing the correlations among the parameters. 
We recall that by construction all the empirical parameters are a-priori uncorrelated to each other.
This means that the possible correlations exhibited by $p_2(P_\alpha,P_\beta)$ will be physical correlations induced by the astrophysical requirements.

The 1-parameter probabilities $p_1(P_\alpha)$ are represented in Fig.~\ref{fig:prob} for the three scenarii DURCA-0 (black), DURCA-1 (red), and DURCA-2 (blue). 
For convenience, the prior distribution is also represented (dashed line).
Comparing $p_1$ to the initial prior distribution is instructive: some parameters are only weakly modified by the filters and the dUrca scenarii, like $E_{sym}$, $L_{sym}$ and $K_{sat}$,
while some other parameters are modified, like $K_{sym}$, $Q_{sat/sym}$ and $Z_{sat/sym}$.
The two main reasons why the parameters $E_{sym}$, $L_{sym}$ and $K_{sat}$ are very weakly impacted by the filters are i) these parameters are already 
well constrained by nuclear physics knowledge (their domain of variation is rather small compared to the others) and ii) the filters probe properties much beyond 
saturation density, where these low order empirical parameters are weakly effective.

It can be observed in Fig.~\ref{fig:prob} that the probabilities associated to the parameters $Q_{sat/sym}$ and  $Z_{sat/sym}$ are slightly more peaked and more narrow
than the prior distribution.
There is also a small but systematical shift between the distributions associated to DURCA-0 on one side and DURCA-1 and 2 on the other side:
the $Q_{sat}$ and $Z_{sat}$ distributions are slightly shifted to the left, making symmetric matter softer, for DURCA-1 and 2 compared to DURCA-0.
However, an opposite trend is observed for the isovector parameters $Q_{sym}$ and $Z_{sym}$ which are systematically shifted slightly to the right 
for DURCA-1 and 2 compared to DURCA-0.
This last shift can be understood from the fact that DURCA-1 and 2 select larger values of the symmetry energy $S(n_0)$ then DURCA-0, since they require
larger proton fraction above saturation density.

Finally, the largest effect observed in Fig.~\ref{fig:prob} is for the empirical parameter $K_{sym}$: the probability distribution $p_{lik}$ is narrower than the prior one,
and the three hypothesis DURCA-0, 1 and 2 produce a systematic shift to the right (increasing values for $K_{sym}$).
From Fig.~\ref{fig:prob} we notice that only $K_{sym}$ is impacted by all of the three hypothesis.
This can be understood from the fact that $K_{sym}$ is the most effective parameter influencing the proton fraction for densities corresponding to the range of masses
between $1.6M_{\odot}$ and $2M_{\odot}$.
Anticipating the following results, this mass range corresponds to central densities between 2 and 3 $n_{sat}$.

\begin{table*}[t]
\centering
\setlength{\tabcolsep}{1.5pt}
\renewcommand{\arraystretch}{1.6}
\begin{ruledtabular}
\begin{tabular}{cccccccccc}
&  & $E_{sym}$ & $L_{sym}$ & $K_{sat}$ & $K_{sym}$ & $Q_{sat}$ & $Q_{sym}$ & $Z_{sat}$ & $Z_{sym}$ \\
 &         & MeV           & MeV           & MeV           & MeV           & MeV           & MeV           & MeV           & MeV \\
\hline
Prior         &            & 32$\pm$2  & 60$\pm$15   & 230$\pm$20   & -100$\pm$100   & 300$\pm$400   & 0$\pm$400      & -500$\pm$1000          & -500$\pm$1000 \\ 
\hline
DURCA-0 & $p_1$  & 31.9$\pm$2.0  & 57.6$\pm$13.5   & 232.5$\pm$18.0   & -103$\pm$76   & 390$\pm$313   & 115$\pm$317      & -424$\pm$883    & -720$\pm$900 \\ 
                 & $p_{1,RM}$  & 32.1$\pm$4.0  & 50.3$\pm$16.8   & 230.8$\pm$28.2   & -132.5$\pm$168   & 1056$\pm$592   & 179$\pm$1056  & -130$\pm$2453   & -288$\pm$3027 \\ 
            \hline
DURCA-1 & $p_1$   & 31.6$\pm$1.9     & 56.6$\pm$12.7    & 231.7$\pm$18.3   & -73$\pm$75    & 267$\pm$321  & 340$\pm$316    & -650$\pm$863      & -389$\pm$902 \\ 
              & $p_{1,RM}$     & 31.7$\pm$3.9     & 49.1$\pm$16.2    & 230.3$\pm$28.2   & -96$\pm$163   & 661$\pm$555  & 960$\pm$839    & -621$\pm$2481  & 598$\pm$2928 \\ 
            \hline
DURCA-2 & $p_1$   & 31.7$\pm$1.9     & 58.5$\pm$12.4    & 231.6$\pm$18.2   & -48$\pm$74       & 256$\pm$319  & 344$\pm$322    & -634$\pm$848    & -500$\pm$937 \\ 
              & $p_{1,RM}$     & 31.8$\pm$3.8     & 50.5$\pm$16.5    & 229.9$\pm$28.2   & -47$\pm$162    &  592$\pm$584  & 985$\pm$833   & -394$\pm$2464   & 440$\pm$3001 \\ 
\end{tabular}
\end{ruledtabular}
\caption{Centroids and standard deviations associated to the 1-parameter probability $p_1(\alpha_1)$ discussed in Sec.~\ref{sec:impactep}; and to the distribution $p_{1,RM}$ discussed in Sec.~\ref{sec:inversion} and corresponding to the TOV inversion problem.}
\label{tab:epcentroids}
%\end{center}
\end{table*}
%\end{sidewaystable}

\begin{figure*}[tb]
\begin{center}
\includegraphics[angle=0,width=0.7\linewidth]{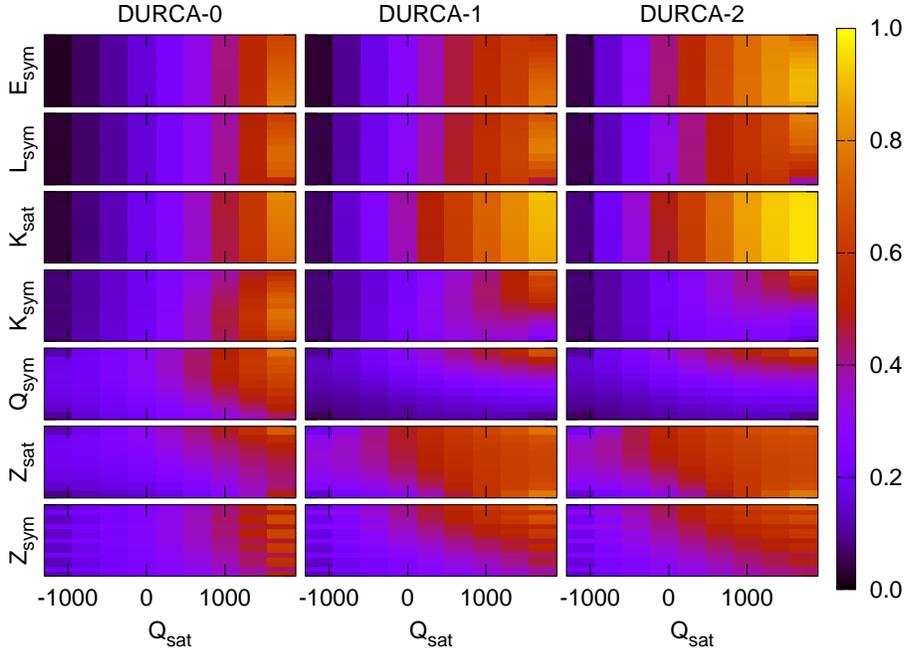}
\end{center}
\caption{(Color online) Two parameters probabilities function of $Q_{sat}$ for DURCA-0, DURCA-1 and DURCA-2 (the prior is here taken flat). The units on the x-axis are MeV and the y-axes cover the range defined in Tab.~\ref{tab:epbound} from bottom to top.}
\label{fig:proba2a} 
\end{figure*}

\begin{figure*}[tb]
\begin{center}
\includegraphics[angle=0,width=0.7\linewidth]{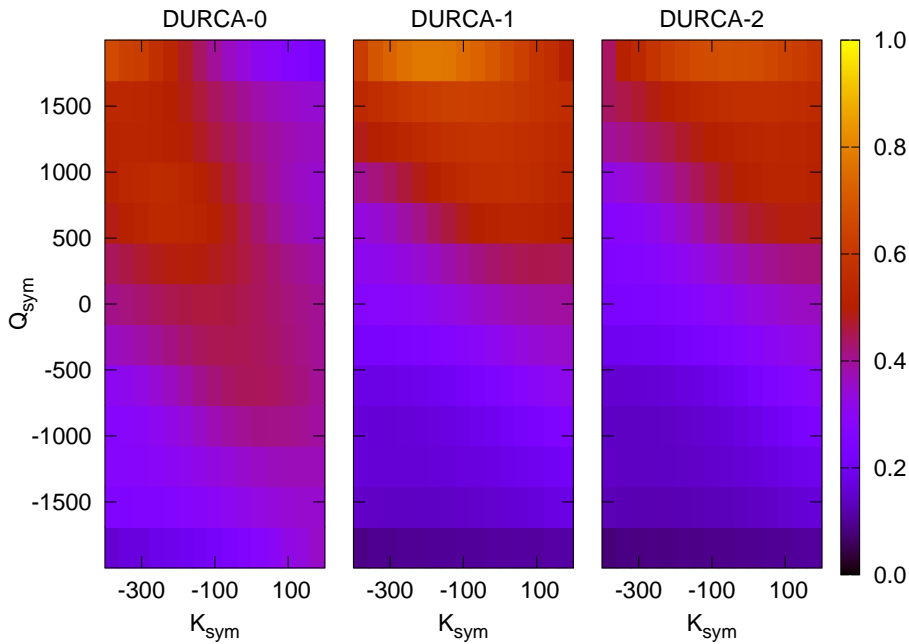}
\end{center}
\caption{(Color online) Two parameters probabilities showing the correlation between $Q_{sym}$ and $K_{sym}$ for DURCA-0, 1 and 2 (the prior is here taken uniform). The units on the axes are MeV.}
\label{fig:proba2b} 
\end{figure*}

From the 1-parameter probability distributions $p_1$, it is possible to define a centroid ($\langle P_\alpha\rangle$) and a standard deviation ($\sigma_\alpha$) as
\begin{eqnarray}
\langle P_\alpha\rangle &=& \int dP_\alpha \, P_\alpha \, p(P_\alpha)  \, ,\\
\sigma_\alpha^2 &=&  \int dP_\alpha \, \left[ P_\alpha - \langle P_\alpha\rangle \right]^2 p(P_\alpha) \, .
\end{eqnarray}
For the 8 parameters represented in Fig.~\ref{fig:prob}, the centroids and the standard deviations deduced from $p_1$ are given in Tab.~\ref{tab:epcentroids} for the three scenarii DURCA-0, 1, and 2.
%The line corresponding to $p_R$ is discussion in a following section.
We recall also the prior distribution on the first line of Tab.~\ref{tab:epcentroids}.
It is interesting to extract the positions of the central values for some empirical parameters.
For instance, the preferred value for $K_{sym}$ is -103~MeV for DURCA-0, -73~MeV for DURCA-1 and -48~MeV for DURCA-2.

The uncertainties remain however quite large for the parameters for which we expected better constraints after the selection of physical constraints.
In other words, the distribution of empirical parameters is not substantially impacted by the filtering.
This can be understood from the fact that the impact of taking an empirical parameter away from the preferred one can be compensated by a change of the other empirical parameters.
Therefore, the effect of the filtering remains weak for the probability distribution $p_1$.
The effect of the compensation phenomenon can be better appreciated in the 2-parameter probability distribution or in the correlation matrix, which we now turn to examine.

The 2-parameters probability distribution $p_2(P_\alpha,P_\beta)$ is interesting since it shows the correlations among empirical parameters which are induced by the filtering conditions. 
We show in Figs.~\ref{fig:proba2a} and \ref{fig:proba2b} some selected 2-parameters probabilities where we have removed the influence of the prior distribution by considering a flat prior: all the empirical parameters are varied evenly between the minimum and maximum values provided in Tab.~\ref{tab:epbound}.
In this way, we can check the effects of the filtering without the influence of the chosen prior probability.

In Fig.~\ref{fig:proba2a}, we show the probability $p_2(Q_{sat},P_\alpha)$ (without prior) for all possible empirical parameters $P_\alpha$. 
It is interesting to note that the effect of the filtering conditions is to systematically prefer large and positive values for $Q_{sat}$.
As we noticed from the analysis of Fig.~\ref{fig:MRsat}, low values of $Q_{sat}$ around 0 and lower soften substantially the EOS, which can lead to $v_s^2<0$, corresponding to an unphysical spinodal instability
at high density.
As seen in Fig.~\ref{fig:proba2a}, the exclusion of such  instability induces a positive preferred value for $Q_{sat}$.
{The high values of $Q_{sat}$ are presently unbounded since we impose only a lower boundary on the maximal mass, $M>2M_\odot$. 
In a future analysis, we may explore the impact of assuming an upper boundary for the maximal mass on the largest values of $Q_{sat}$.}
We can also see that the posterior probabilities are very similar 
independent of the adopted condition for dUrca.

In Fig.~\ref{fig:proba2b}, we represent the 2-parameters probability distribution $p_2(K_{sym},Q_{sym})$ (without prior) for DURCA-0, 1 and 2.
A negative correlation can be observed between $K_{sym}$ and $Q_{sym})$, especially for DURCA-0: the larger $K_{sym}$ the lower $Q_{sym})$.
It illustrates the impact of the condition on the proton fraction.
For DURCA-0, since the proton fraction should remain small up to a density corresponding to $2M_\odot$, a increase of $K_{sym}$ which would violate this constraint is compensated by a decrease of $Q_{sym}$.
For DURCA-1 and DURCA-2, the correlation is weaker since both $K_{sym}$ and $Q_{sym}$ can be large and still  satisfy the condition on the proton fraction.
However, if $K_{sym}$ is too large, it may also induce a supra-luminal EOS which is forbidden up to a density corresponding to $2M_\odot$.
These two conditions can be viewed in the correlation pattern shown in Fig.~\ref{fig:proba2b}.

A more clear and compact way to represent the correlations is to evaluate the correlation matrix.
In addition, the correlation matrix provides a quantitative measure of the strength of the correlation.
The correlation matrix is defined as
\begin{eqnarray}
\mathrm{corr}(P_\alpha,P_\beta) = \frac{ \mathrm{cov}(P_\alpha,P_\beta) }{\sigma_{\alpha} \sigma_{\beta}} \, ,
\end{eqnarray}
where the covariance matrix is defined as
\begin{eqnarray}
\mathrm{cov}(P_\alpha,P_\beta) 
&=&\!\! \int\!\! dP_\alpha \int\!\! dP_\beta \, \big[ P_{\alpha} - \langle P_{\alpha}\rangle \big] \big[ P_{\beta} - \langle P_{\beta}\rangle \big] p(P_\alpha,P_\beta) \, . \nonumber \\
\end{eqnarray}

The correlation matrices $\mathrm{corr}(P_\alpha,P_\beta)$ for the 8 empirical parameters and for the three scenarii DURCA-0, 1, and 2 are shown in Fig.~\ref{fig:correlationmatrix}.
We remind that the prior distribution is diagonal in the empirical parameters, and therefore it cannot generate non-diagonal matrix elements in the correlation matrix.
The non-diagonal matrix elements can only be generated from the physical conditions.
This is at variance with popular EOS modelings, such as for instance Skyrme functionals, 
where the functional form of the energy density is such that it may generate a-priori correlations among the empirical parameters which might have no physical meaning.

The correlation matrix takes values close to zero if the (linear) correlations between the parameters $P_\alpha$ and $P_\beta$ are very weak, it approaches 1 for strong correlations and -1 for strong anti-correlation.
As a rule of thumb, $|c|<0.5$ denotes a negligible correlation, and one cannot speak of a strong correlation unless $|c|>0.8$.

\begin{figure}[tb]
\begin{center}
\includegraphics[angle=0,width=0.99\linewidth]{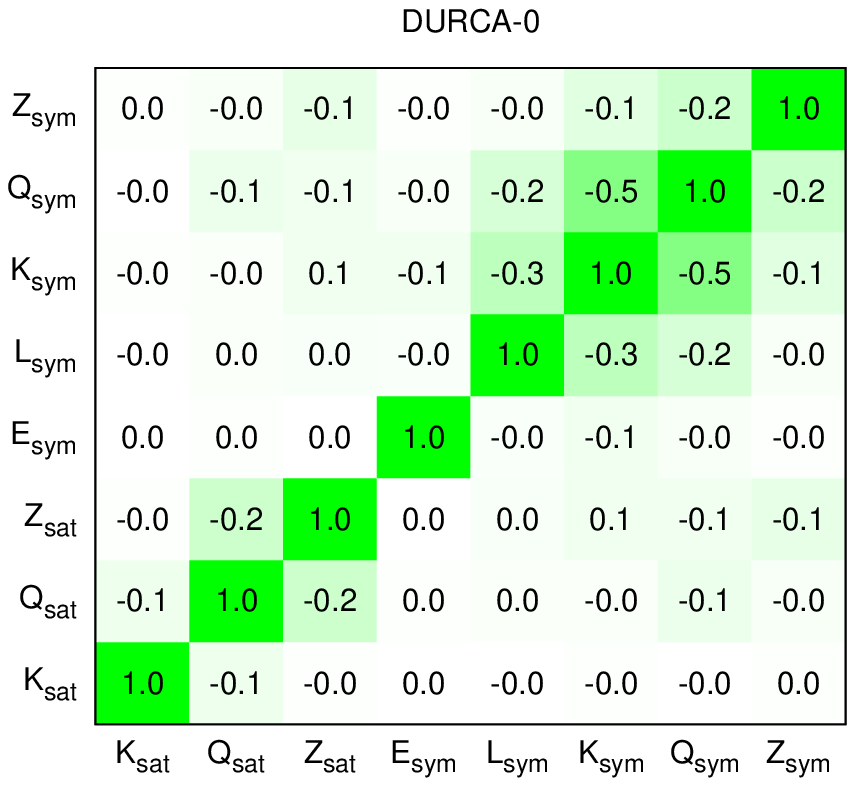}
\includegraphics[angle=0,width=0.99\linewidth]{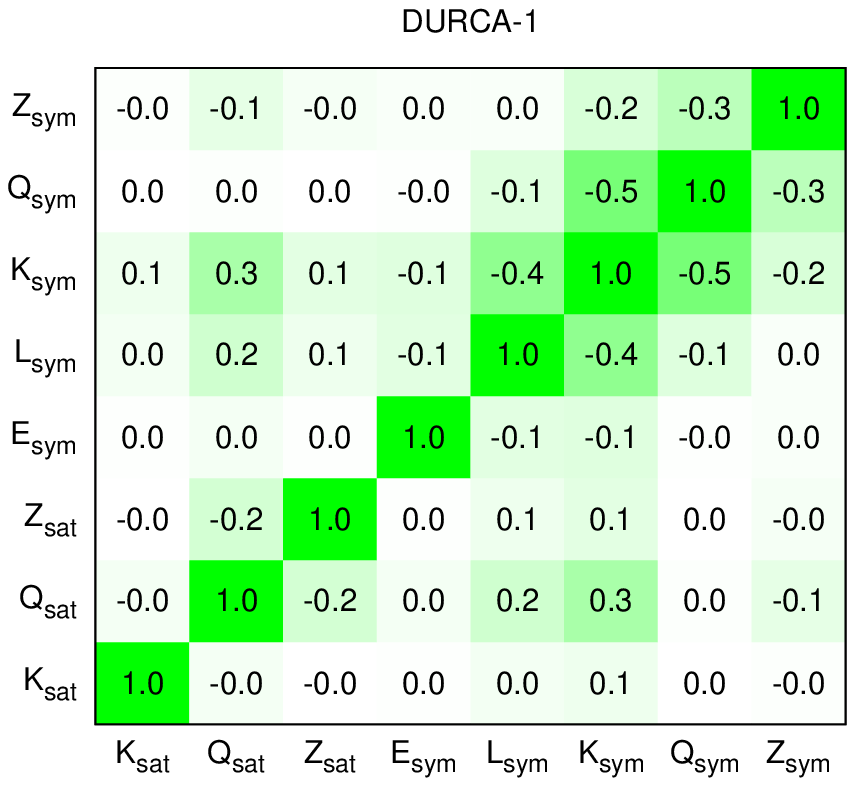}
\includegraphics[angle=0,width=0.99\linewidth]{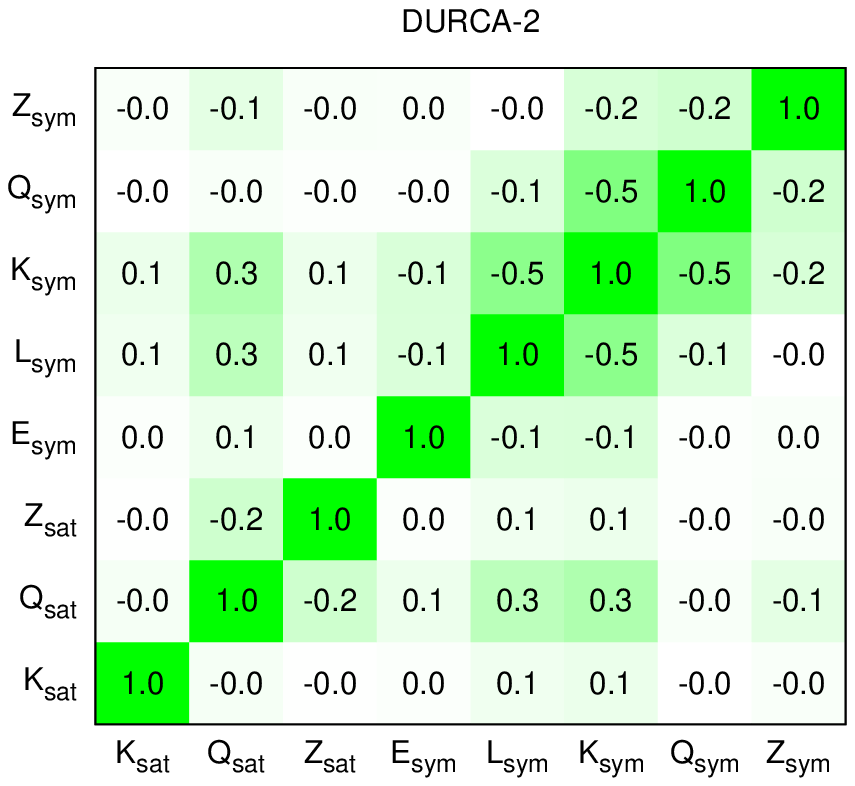}
\end{center}
\caption{(Color online) Correlation matrix for the 8 empirical parameters and the three scenarii DURCA-0, 1 and 2. The color index goes from 0 (white) to 1 in absolute value (green) in a linear scale as shown on the right bars shown on each graphs. See text for discussion.}
\label{fig:correlationmatrix} 
\end{figure}

We first remark from Fig.~\ref{fig:correlationmatrix} that the matrix correlation exhibits weak correlations in general.
It is interesting to compare such matrix correlation to similar ones generated from Skyrme density functional~\cite{Kortelainen2010,McDonnell2015,Rocamaza2015}.
Other approaches comparing a large variety of different methods, see Ref.~\cite{Klahn2006,Dutra2012} for instance, have also concluded that filtering among physical EOS 
considerably reduces the number of EOS.
Performing such a comparison is not straightforward since the physical filters are not exactly the same in our present work and in these papers.
In our case, our selection filter is much less constraining than in the other works.
However, the flexibility of the EOS considered in these other works is  lower than in our present study.
This flexibility is clearly an advantage which is at the origin of our approach, suppressing spurious correlations among empirical parameters, see paper~I.
This flexibility, i.e. the absence of a-priori correlations among the empirical parameters, is the reason why the width of $p_1$ remains large after applying the physical requirements, and  the correlation
matrix present weak off-diagonal matrix elements.

Some matrix elements shown in  Fig.~\ref{fig:correlationmatrix} depart from zero and even if they remain weak, they are large enough to be noticed.
They indicate the existence of correlations among empirical parameters induced by the physical constraints. 
These correlations are of two kinds: there are correlations between only two parameters, and there are block correlations.
The block correlations reveals the existence of complex multi-parameter correlations.
We shall also notice that some are correlations (positive matrix elements) and some are anti-correlations (negative matrix elements).
Let us now describe these correlations in more detail.

We notice a large block of weak correlations among the isovector empirical parameters. 
They are induced by the hypothesis made for the dUrca process and do not vary much in strength between one hypothesis to another. 
A slight shift towards lower order empirical parameters can however be noticed as one compare DURCA-0, 1 and 2.
This is not surprising since the hypothesis DURCA-2 is the one which give the strongest constraint at low density (or mass) while DURCA-0 constrains higher densities (or masses).
In addition these correlations are negative, indicating compensation effects between these empirical parameters.

It is interesting to note some weak single parameters correlations.
The first one is a very slight anti-correlation between $Q_{sat}$ and $Z_{sat}$ which is related to the causality constraint.
The second ones are weak positive correlations between $Q_{sat}$ and $K_{sym}$ which appear only for the DURCA-1 and 2 hypothesis. 
They are absent for DURCA-0 {and reveal positive correlations between IS and IV channels induced by the dUrca condition.
It shows a tendency for the EOS satisfying DURCA-1 and 2 to be slightly more repulsive than the ones satisfying DURCA-0.
This tendency is weak but could be observed, as we will see in the following.}
These correlations reveal {a weak, but still understandable, correlation between} the dUrca hypothesis on the empirical parameters.

In conclusion, the correlation matrix reveals interesting but weak multi-parameter correlations generated by the causality condition and the hypothesis for the dUrca process. 
These correlations remain however extremely weak.
This is an interesting observation, showing that general constraints related to causality and dUrca process are not very influential on the {correlation among the empirical parameters}.
If such constraints play an important role in a given modeling, this might be more related to the lack of flexibility of the considered EOS, more than on the physical effect of the constraints.
This analysis also shows the interest of the correlation matrix analysis within our approach, which should be further explored with additional constraints in the future such that for instance the masses and radii of finite nuclei~\cite{Chatterjee2017}.

\begin{figure*}[t]
\begin{center}
\includegraphics[angle=0,width=0.8\linewidth]{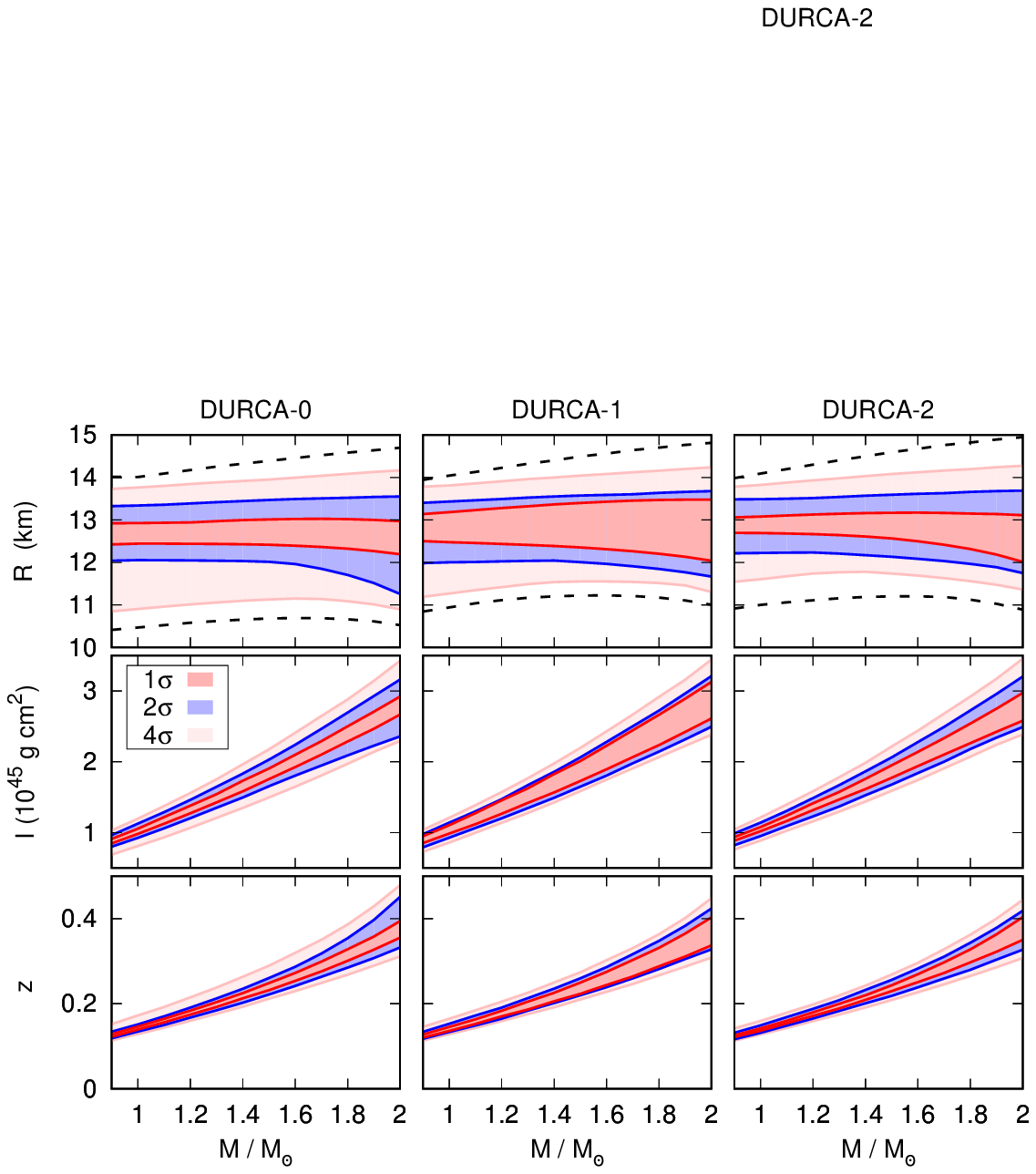}
\end{center}
\caption{(Color online) $1\sigma$-CL, $2\sigma$-CL and $4\sigma$-CL domains for the correlation between the total radius $R$, moment of inertia $I$ and surface redshift $z$ function of the mass $M$ for DURCA-0, 1, and 2 scenarii. The dashed lines on the top panels represent the maximum and minimal values of the radius. See text for discussion.}
\label{fig:impactMR1} 
\end{figure*}

\subsection{Impact on the global properties of neutron stars}
\label{sec:impactns}

In this section, we continue with the statistical analysis of the EOS and analyze the predictions  
of the global properties of NS. 
We consider the same prior and likelihood probabilities as the ones defined in Sec.~\ref{sec:impactep}, see Eqs.~(\ref{eq:probalikely})-(\ref{eq:probaprior}).

To better quantify our results, 
we generate confidence level (CL) domains for the different observables.
In practice, we run over the 25 millions of meta-EOS and group them based on their likelihood probability $p_{lik}$, see Eq.~(\ref{eq:probalikely}).
The group for which the probability $p_{lik}\ge p_{lik, max}e^{-1/2}$, corresponds to $1\sigma$-CL around the maximum value of the probability, $p_{lik, max}$.
Similarly, the $2\sigma$-CL corresponds to $p_{lik}\ge p_{lik, max}e^{-2}$, the $3\sigma$-CL to $p_{lik}\ge p_{lik, max}e^{-9/2}$ and the $4\sigma$-CL to 
$p_{lik}\ge p_{lik, max}e^{-8}$.
For a Gaussian probability distribution, $1\sigma$-CL represents about 67\% of the data around the best probability, $2\sigma$-CL about 95\%, 
$3\sigma$-CL about 99.9\%, and $4\sigma$-CL almost 100\%.

From the 25 millions of initial meta-EOS, we finally find about 16 meta-EOS in the $1\sigma$-CL group for DURCA-0 hypothesis (10 for DURCA-1 and 5 for DURCA-2),
650 meta-EOS  in the $2\sigma$-CL group for DURCA-0 hypothesis (160 for DURCA-1 and 140 for DURCA-2), and
75,000 meta-EOS  in the $4\sigma$-CL group for DURCA-0 hypothesis (12,000 for DURCA-1 and 14,000 for DURCA-2).

We transform the likelihood probability in terms of the parameters $p_{lik}(\{P_\alpha\}_{i_{n\sigma}} )$ into a probability distribution function of the NS global properties, such as its mass and radius, according to the following transformation,
\begin{equation}
p_{MR}^{n\sigma}(M,R) = \sum_{i\in{n\sigma}-CL} p_{lik}(\{P_\alpha\}_{i} ) \delta(M_\alpha-M) \delta(R_\alpha-R),
\end{equation}
where $M_\alpha$ and $R_\alpha$ run over the solution of the TOV equation for a given parameter set  $\{P_\alpha\}_{i}$. 
In practice, masses (radii) are grouped into 13 (200) bins according to the following algorithm,
\begin{eqnarray}
M(k_M)&=&M_{min}+(k_M-1) \Delta M \, , \\
R(k_R) &=&R_{min}+(k_R-1) \Delta R \, ,
\end{eqnarray}
where $M_{min}=0.8M_\odot$, $\Delta M=0.1M_\odot$, $R_{min}=9.5$~km, $\Delta R=50$~m, and 
the indexes $k_M=1,\dots,13$ and $k_R=1,\dots,200$.

\begin{figure*}[t]
\begin{center}
\includegraphics[angle=0,width=0.8\linewidth]{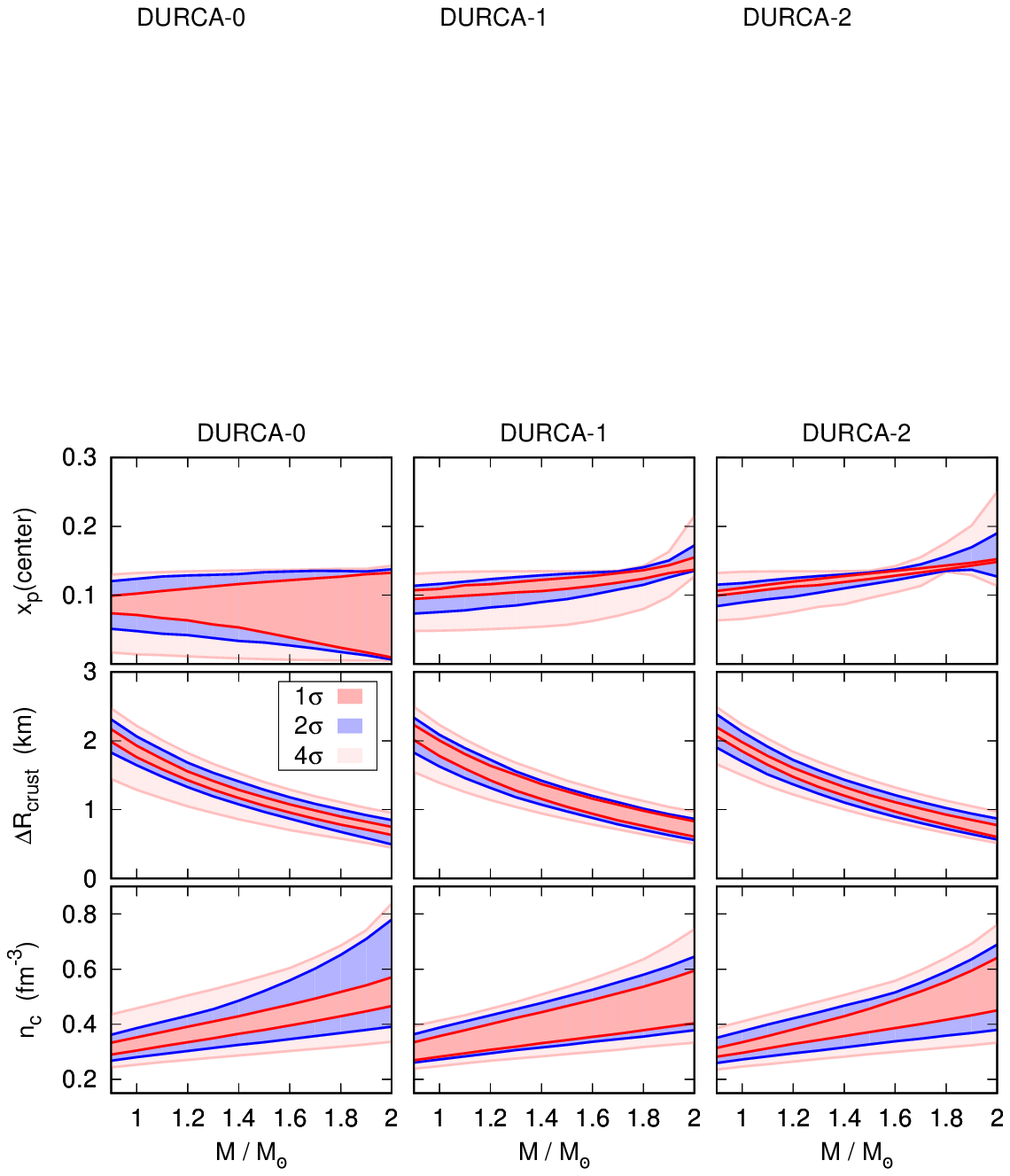}
\end{center}
\caption{(Color online) Same as Fig.~\ref{fig:impactMR1} for the correlation between the central proton fraction $x_p$, crust thickness $\Delta R_{crust}$ and central density $n_c$ function of the mass.}
\label{fig:impactMR2} 
\end{figure*}

Similarly one can define the probabilities $p_{IR}^{n\sigma}(I,R)$ between the moment of inertia $I$ and the mass $M$, 
$p_{zR}^{n\sigma}(z,R)$ between the surface redshift $z$ and the mass,
$p_{x_pR}^{n\sigma}(x_p,R)$ between the proton fraction at the center and the mass,
$p_{\Delta R_{crust}R}^{n\sigma}(\Delta R_{crust},R)$ between the crust thickness and the mass,
and $p_{n_cR}^{n\sigma}(n_c,R)$ between the central density and the mass.

All these probabilities are shown in Figs.~\ref{fig:impactMR1} and \ref{fig:impactMR2} for the different hypothesis DURCA-0, 1, and 2.
The inner domain (in red) is the $1\sigma$-CL, then comes the $2\sigma$-CL (in blue) and the $4\sigma$-CL (in pink).
Let us first comment Fig.~\ref{fig:impactMR1} where are shown some very general properties of NS such as their radii,
the moment of inertia, and their surface redshift.
Despite a rather general agreement among the predictions based on the three hypothesis DURCA-0, 1 and 2, one can notice some differences: 
The upper bound for the radius function of the mass is rather independent of the hypothesis, but the lower band shows some small differences.
Specifically, the DURCA-0 hypothesis allows smaller radii %, especially at high mass, 
compared to the two other hypothesis.
More precisely, the smallest radius is about 10.5~km for DURCA-0, while it is about 11~km for DURCA-1 and 2.
The average radius is almost independent of the mass and of the dUrca hypothesis.
It is evaluated to be between 12 and 13.5~km for the 1-$\sigma$ contour. The 2-$\sigma$ contour is sligthly larger, and allow lower radii at high mass for the DURCA-0 hypothesis.
More accurate estimation will be given further in our analysis, see Fig.~\ref{fig:impactMR3}.
%approximately 12.7$\pm0.4$~km at 1$\sigma$.
%for DURCA-1 and 2, while it decreases from about 12.8$\pm$0.3~km for 
%$M=0.8M_\odot$ down to about 12$\pm$0.6~km for $M=2M_\odot$.
%As a consequence, the moment of inertia can be slightly smaller for DURCA-0, and the redshift $z$ can be slightly larger.
In addition, the observation of a NS with a radius between 10 and 11~km NS would be incompatible with DURCA-1 and 2 hypothesis, but will still be marginally 
compatible with the DURCA-0 hypothesis.

In Fig.~ \ref{fig:impactMR2} we analyze internal properties of NS such as the central proton fraction $x_p($center$)$, the thickness of the crust (inner+outer crust)
$\Delta R_{crust}$ and the central density $n_c$.
As expected, there is a clear difference in the central proton fraction predicted by the dUrca hypothesis: DURCA-0 favors low values of the proton fraction 
(below 1/9 in the whole density domain), DURCA-1 favors values which can be above about 1/9\%, and DURCA-2 even slightly larger proton fractions.
%While there is almost no impact on the crust thickness,
%the central density shows however some differences. The EOS satisfying the DURCA-0 hypothesis are generally rather soft, and as a consequence, many of these EOS predict
%larger central densities compared to DURCA-1 and 2 hypothesis at the same mass.
%This tendency is clearly visible for NS which are at $2M_\odot$.
There is almost no impact of the different proton fraction on the crust thickness and central density.

In summary, Figs.~\ref{fig:impactMR1} and \ref{fig:impactMR2} show that global properties of NS are rather universal and weakly influenced by the dUrca hypothesis.
%One could however remark some differences among these predictions.
%Some properties like the crust thickness are almost universal, some other properties like the moment of inertia and the redshift are only weakly influenced, 
%and finally, properties like the central proton fraction, the central density and the radius show a larger impact.
%The radius tends to decrease as a function of the mass for DURCA-0, while it is almost independent of the mass for DURCA-1 and 2.
As already proposed in Ref.~\cite{Blaschke2016}, there is an interesting universality of the EOS under the condition of charge neutrality and $\beta$-equilibrium.
While in Ref.~\cite{Blaschke2016} the authors suggested that the universality behavior holds only for the EOS which prevent dUrca (corresponding to the DURCA-0 
hypothesis in our work), we generalize this universal behavior to EOS which allow dUrca process for high mass NS (DURCA-1 and 2). 

\begin{figure}[tb]
\begin{center}
\includegraphics[angle=0,width=1.0\linewidth]{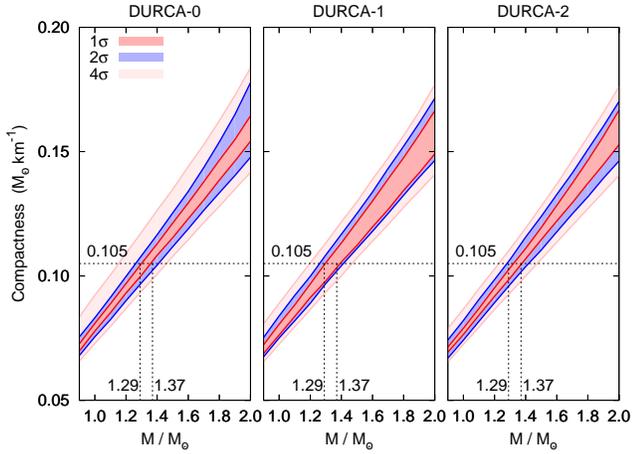}
\end{center}
\caption{(Color online) Compactness $(M/M_\odot)/(R/km)$ as a function of the mass for the three hypothesis DURCA-0, 1, 2.}
\label{fig:CP} 
\end{figure}

\begin{figure*}[t]
\begin{center}
\includegraphics[angle=0,width=0.7\linewidth]{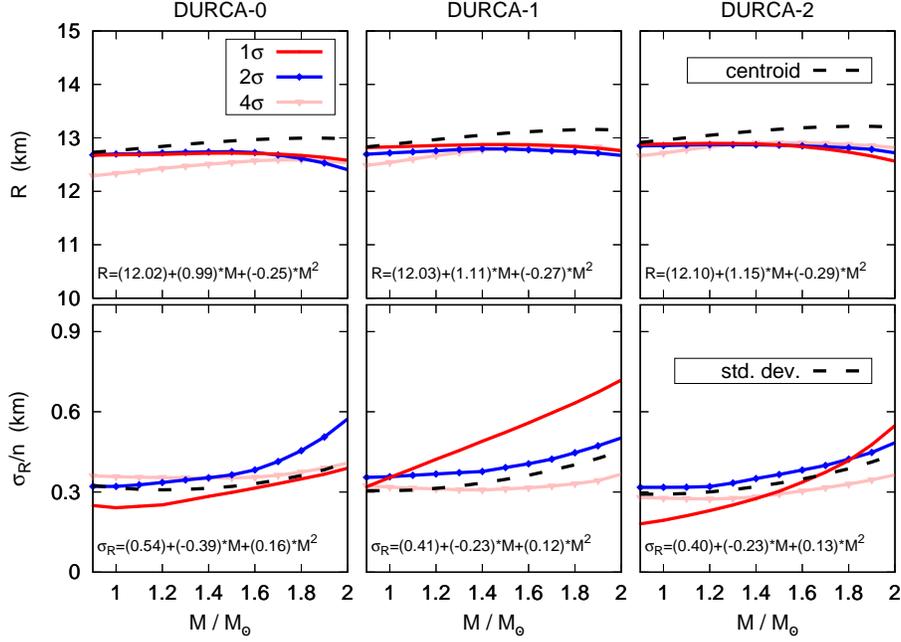}
\end{center}
\caption{(Color online) Centroids and standard deviation of the NS radii function of the mass, with DURCA-0, 1, 2.}
\label{fig:impactMR3} 
\end{figure*}

We show in Fig.~\ref{fig:CP} the compactness defined as $(M/M_\odot)/(R/km)$ where $R$ is expressed in km for the three
hypothesis DURCA-0, 1, 2.
The influence of dUrca hypothesis is very weak, and 
the M-dependence of the compactness appear to be universal here also. %, especially for masses $M<1.8M_\odot$. 
%The only effect is that DURCA-0 predicts slightly larger values of the compactness for large masses. 
{It is therefore interesting to note the stability of the relation between the compactness and the mass (independent of the dUrca hypothesis), especially for low values of the compactness ($<0.12$) where it can safely be assumed that matter is composed of nucleons.}

{It was recently claimed that the compactness of the isolated NS RX J0720.4-3125 is 0.105$\pm$0.002~\cite{Hambaryan2017}.
It is interesting to illustrate the use of the correlation between the compactness and the mass to infer the mass from this extremely accurate estimation of the compactness.
The value for the compactness is reported in Fig.~\ref{fig:CP} and since it appears to be in the domain where the compactness is quite stable, we can use the $1\sigma$-CL to estimate
the mass of RX J0720.4-3125.}
The construction lines are shown in Fig.~\ref{fig:CP} and we predict that RX J0720.4-3125 has a mass of 1.33$\pm$0.04~M$_\odot$ at the $1\sigma$ level.
The only hypothesis we have made concerning the EOS is that it is nucleonic and respects minimal physical constraints.

From Fig.~\ref{fig:impactMR2}, we deduce that the 1.33$M_\odot$ NS have central densities less than 2-2.5$n_{sat}$.
Since most of the EOS predicting phase transition to hyperon or quark matter always predict it to be above about 3$n_{sat}$, we can conclude 
that our hypothesis of nucleonic matter for RX J0720.4-3125 is rather safe, and therefore our predicted mass is quite realistic.
From Fig.~\ref{fig:impactMR1}, we can also predict that the radius of RX J0720.4-3125 is 12.7$\pm$0.3~km at the $1\sigma$ level.

\begin{figure*}[p]
\begin{center}
\includegraphics[angle=0,width=0.7\linewidth]{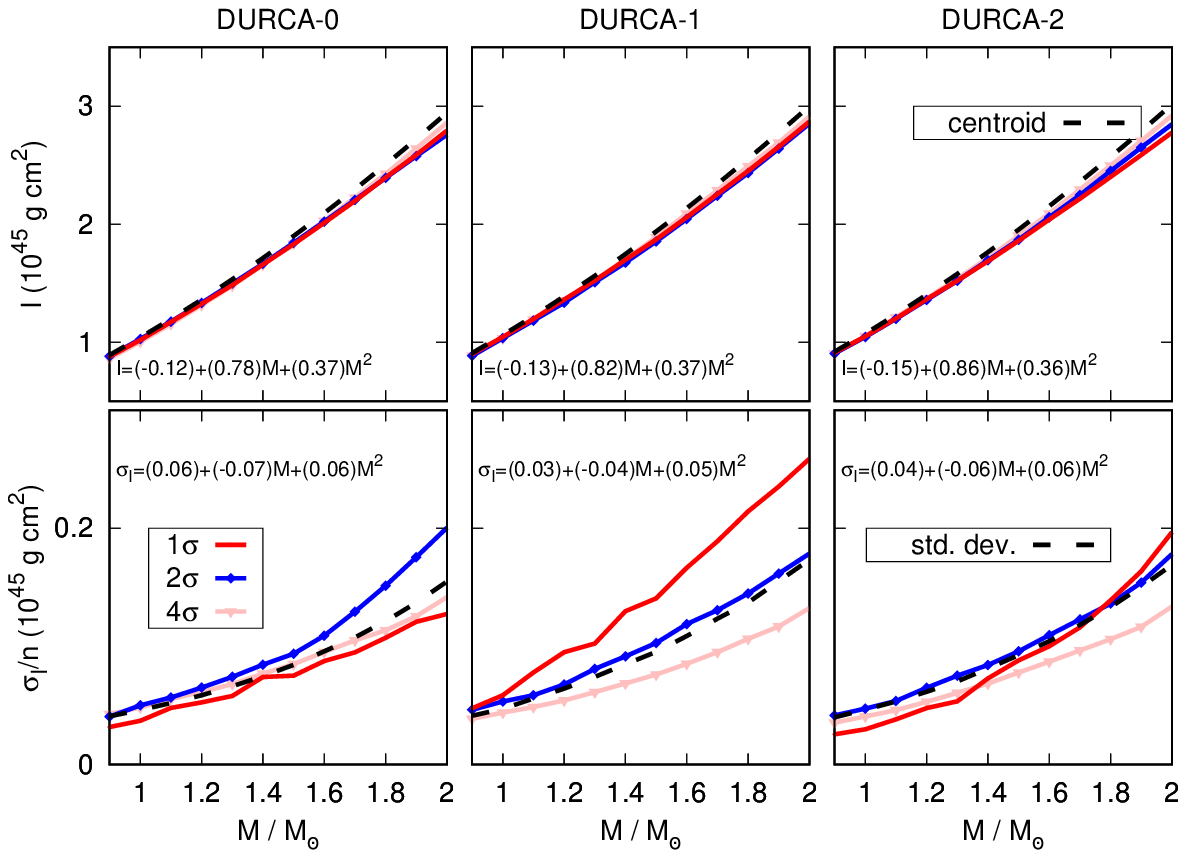}
\end{center}
\caption{(Color online) Same as Fig.~\ref{fig:impactMR3} for the moment of inertia.}
\label{fig:impactMR4} 
\end{figure*}

\begin{figure*}[p]
\begin{center}
\includegraphics[angle=0,width=0.7\linewidth]{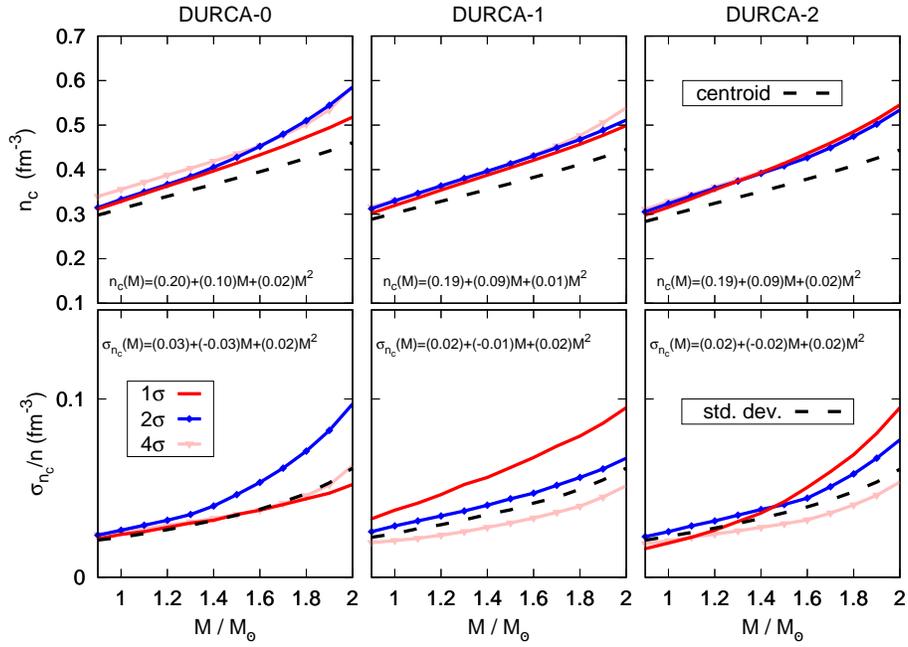}
\end{center}
\caption{(Color online) Same as Fig.~\ref{fig:impactMR3} for the central density.}
\label{fig:impactMR5} 
\end{figure*}

%\begin{figure*}[tb]
%\begin{center}
%\includegraphics[angle=0,width=0.7\linewidth]{plot-mr-2-DURCA-CRST.eps}
%\end{center}
%\caption{(Color online) Same as Fig.~\ref{fig:impactMR3} for the total crust thickness.}
%\label{fig:impactMR6} 
%\end{figure*}

\begin{figure*}[tb]
\begin{center}
\includegraphics[angle=0,width=0.7\linewidth]{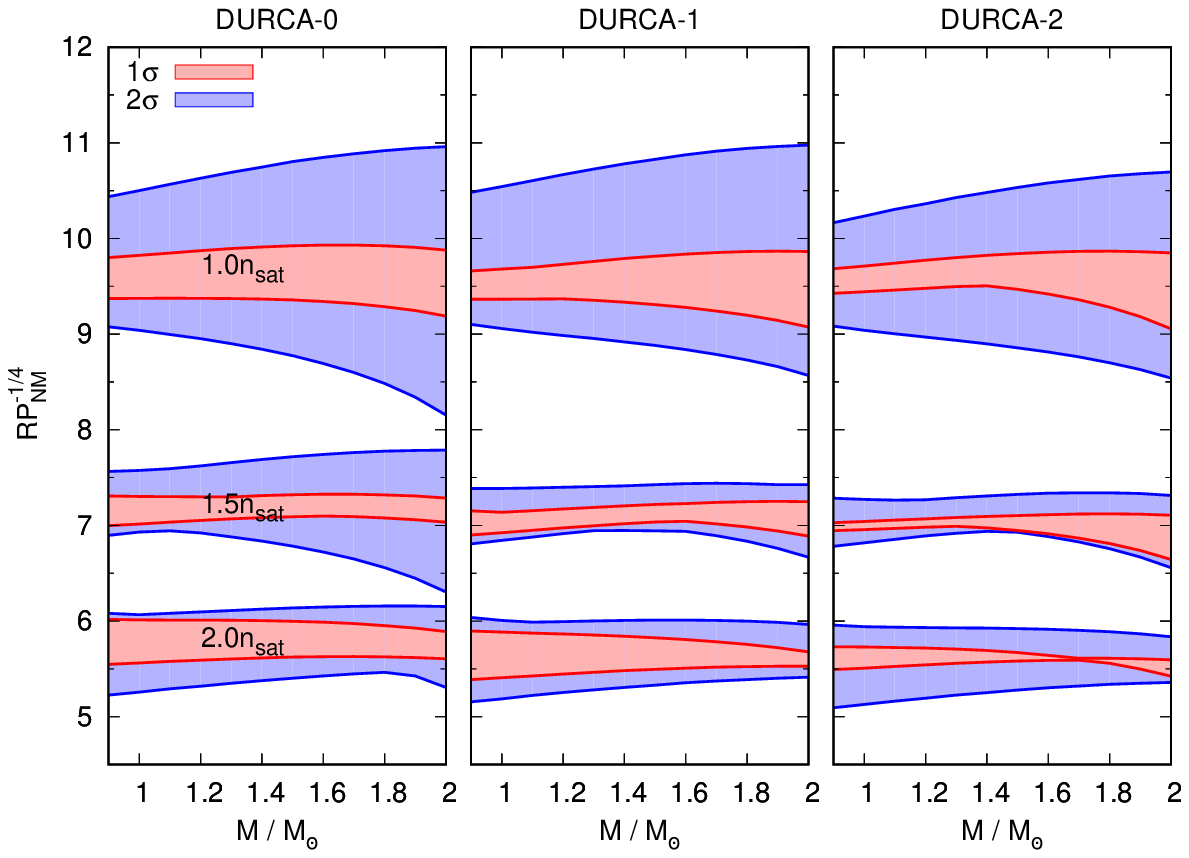}
\end{center}
\caption{(Color online) Empirical relation between the pressure (in MeV~fm$^{-3}$) and the radius (in km) function of the mass.
The pressure is calculated in NM and is defined at $1.0n_{sat}$, $1.5n_{sat}$, or $2.0n_{sat}$ as indicated in the plots.
We used the same units on the y-axis as in Ref.~\cite{Lattimer2001a}.}
\label{fig:impactPR} 
\end{figure*}

Let us now discuss in more details the contours shown in Figs.~\ref{fig:impactMR1} and \ref{fig:impactMR2}. 
In the following, we provide fits to the mass dependence of some NS global properties, including centroids and standard deviation. 
In Fig.~\ref{fig:impactMR3}, the average position of the radius shown in Fig.~\ref{fig:impactMR1} (top panels) are represented.
This radius is calculated in different ways: the first and simplest way is obtained from the mean value between the upper and the lower band
for the radius determined for each $n\sigma$-CL ($n$=1, 2, 4). They are given by the red ($1\sigma$), blue ($2\sigma$), and pink ($4\sigma$)
lines. 
On the bottom panels, the widths of the distributions are calculated from the half difference between the upper and the lower band for the radius.
The widths are divided by $n$ to estimate how close to a Gaussian are the probability distributions.
With dashed lines, we represent the same quantity calculated in a different way:
From the probability $p_{MR}^{n\sigma}(M,R)$ the centroid and the standard deviation are extracted as
\begin{eqnarray}
\langle R_{n\sigma}(M) \rangle^2 &=&  \frac{1}{N_{MR}} \int  R^2 p_{MR}^{n\sigma}(M,R) \,d^3R , \\
\sigma_{R,n\sigma}(M)^2 &=& \frac{1}{N_{MR}} \int  \left[R-\langle R_{n\sigma}(M) \rangle\right]^2 p_{MR}^{n\sigma}(M,R) \,d^3R ,\nonumber \\
\end{eqnarray}
where $N_{MR}=\int d^3R \, p_{MR}^{n\sigma}(M,R)$.
The centroids $\langle R_{n\sigma}(M) \rangle$ (and standard deviation $\sigma_{R,n\sigma}(M)$) are shown in dashed lines on the top (bottom) part in 
Fig.~\ref{fig:impactMR3} considering all the meta-EOS which are inside the $4\sigma$-CL.
Despite some slight differences, the agreement between the different predictions for the average $M$-dependence of the NS radius are all compatible,
within the standard deviation band.
In addition, the standard deviations divided by $n$, where $n$ refers to the order of the CL, are also very similar showing that the distribution of probability
 $p_{MR}^{n\sigma}(M,R)$ is not far from a Gaussian distribution, as far as the first two moments are concerned.
On the bottom part of each panel is provided a second-order in $M$ fit of the centroids and standard deviations.
{From Fig.~\ref{fig:impactMR3} we can conclude that NS, if they are composed exclusively of nucleons and leptons, have a radius of $12.7\pm0.4$~km at 
the $1\sigma$ confidence level. 
%For NS with higher masses (1.6-2$M_\odot$) the uncertainty increases and we predict radii of $12.2\pm0.8$~km at the $1\sigma$ confidence level.}

It is interesting to compare our prediction to other analyzes.
Combining chiral EFT modeling of neutron matter, piecewise polytropes and observed NS masses, 
NS radii have been predicted to range from 10.5 to 13.3 km~\cite{Hebeler2010}.
These boundaries are defined as the maximum and minimum values for the radius, while average value and its dispersion have not been calculated.
Large NS radii, such as 15~km for instance, are excluded as in our analysis.
From a sensibly similar approach, 1.4$M_\odot$ NS radius has been predicted to be in the range 9.7-13.9~km with central densities up to 4.4$n_{sat}$~\cite{Hebeler2013}.
Still based on piecewise polytropes but including observations of both transiently accreting and bursting NS, the radius of a $1.4M_{\odot}$ NS was shown to lie
between 10.4 and 12.9~km in Ref.~\cite{Steiner2013} 
{and between 10.1 and 11.1~km in Ref.~\cite{Ozel2016}.
These two analyzes assume different hypothesis for the photospheric radius expansion mechanism in the analyzes of the burst.
The prediction of Ref.~\cite{Ozel2016}, if confirmed, is difficult to reconcile with the hypothesis that matter only composed of nucleons as in our case.
The other predictions for the radii are more compatible with nuclear matter.
It is not surprising that our estimate for the NS radius lies inside all these boundaries (except those of Ref.~\cite{Ozel2016}) since the considered EOS are more general than only the nuclear EOS, as in our case.}
We obtain smaller uncertainty in our analysis because i) we consider only nuclear EOS, and ii) we consider only empirical parameters compatible with nuclear data analysis.
Recently, the radius of 1.4$M_\odot$ NS was estimated to be in the range 11.09-12.86~km, based on Skyrme EOS~\cite{Alam2016}.
From our analysis, we i) confirm that large NS radii (larger than 14~km) are not compatible with nucleon EOS, ii) and we predict that radii smaller than 11~km are not neither, iii) we state that this uncertainty interval should be associated to any purely nucleonic EOS compatible with empirical constraints, not necessarily Skyrme EOS, and finally iv) any progress in reducing the uncertainties in the critical empirical parameters ($L_{sym}$, $K_{sym}$, $Q_{sat/sym}$) will lead to a reduction of our uncertainty for the NS radius.
If NS radii are ever observed outside our prediction range, then this would be a strong argument in favor of exotic matter EOS.

We have performed a similar analysis for the moment of inertia $I$ in Fig.~\ref{fig:impactMR4} and
and for the central density $n_c$ in Fig.~\ref{fig:impactMR5}.
%for the crust thickness $\Delta R_{crust}$ in Fig.~\ref{fig:impactMR5}.
We conclude from these figures and the probability distributions associated to each of these NS global properties are 
not far from Gaussian up to the second moment
and we provide as well a second-order in $M$ fit of the centroids and standard deviations for these properties.

For the crust thickness we obtained the following fit as function of the mass, where $\Delta R$ and $\sigma_{\Delta R}$ are expressed in km:
\begin{eqnarray}
\Delta R(M) &=& 4.19-2.96M/M_\odot+0.63(M/M_\odot)^2,\\
\sigma_{\Delta R}(M) &=& 0.27-0.22M/M_\odot+0.06(M//M_\odot)^2,
\end{eqnarray}
for DURCA-0,
\begin{eqnarray}
\Delta R(M) &=& 4.23-2.98M/M_\odot+0.63(M/M_\odot)^2,\\
\sigma_{\Delta R}(M) &=& 0.22-0.16M/M_\odot+0.05(M//M_\odot)^2,
\end{eqnarray}
for DURCA-1,
\begin{eqnarray}
\Delta R(M) &=& 4.28-3.01M/M_\odot+0.64(M/M_\odot)^2,\\
\sigma_{\Delta R}(M) &=& 0.21-0.16M/M_\odot+0.04(M//M_\odot)^2,
\end{eqnarray}
for DURCA-2.

{Finally, we test the empirical relation between the pressure and the radius of neutron stars proposed in Ref.~\cite{Lattimer2001a}.
This empirical relation is shown in Fig.~\ref{fig:impactPR} where the pressure is the pressure of neutron matter (NM) calculated at $1.0n_{sat}$, $1.5n_{sat}$, and $2.0n_{sat}$.
We confirm the results obtained in Ref.~\cite{Lattimer2001a}: the spreading among different model is minimized if the pressure is defined at $1.5n_{sat}$, or $2.0n_{sat}$,
in these two cases, the empirical relation is almost independent of the mass of the NS for masses below $1.6M_\odot$.
The values obtained for the empirical relation are also compatible with the ones in Ref.~\cite{Lattimer2001a}.
It should however be noted that we used the pressure in neutron matter in our case while it is the pressure of matter at $\beta$-equilibrium which was used in Ref.~\cite{Lattimer2001a}.}

\subsection{Impact on the equation of state at $\beta$-equilibrium}
\label{sec:impacteos}

\begin{figure*}[tb]
\begin{center}
\includegraphics[angle=0,width=0.7\linewidth]{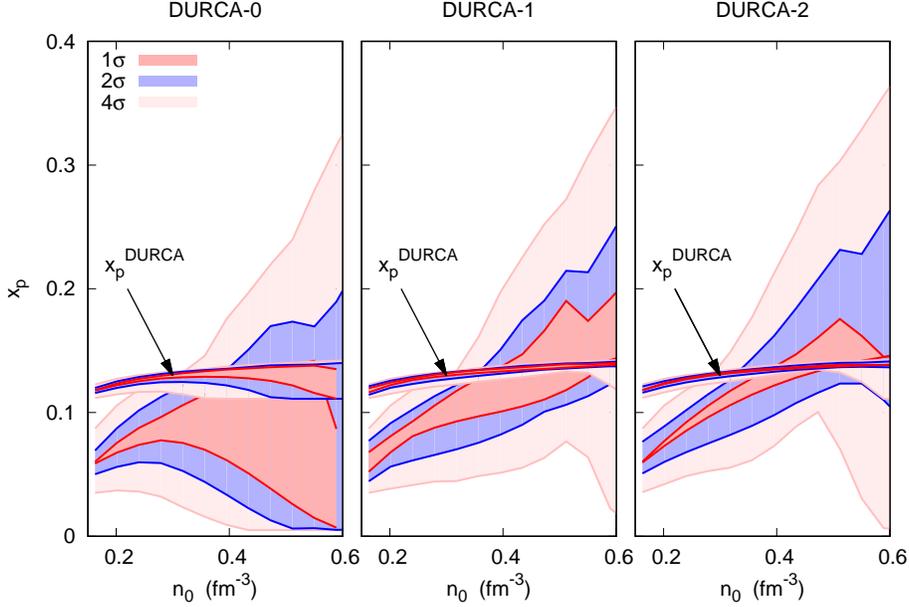}
\end{center}
\caption{(Color online) Impact of the DURCA-0, 1, and 2 scenarii on the proton fraction density dependence. The dUrca threshold condition is also shown. See text for more details}
\label{fig:xp} 
\end{figure*}

We now carry on a similar statistical analysis, but instead of focusing on the global properties of NS, we analyze the  distribution of meta-EOS properties, such as its energy density,
pressure, sound velocity, or distribution of proton fraction as a function of the density $n_0$.
It is interesting to convert the impact of the physical constraints which have been expressed as a function of the NS mass, into the behavior of the EOS properties as a function of the density $n_0$.
There is indeed a strong correlation between the mass and the central density, as shown in Fig.~\ref{fig:impactMR2}, but there is also a non-negligible dispersion of 
this correlation, especially for the large masses.
As a consequence, there is  no one-to-one correspondence between the masses of NS and their central density, and it is interesting to visualize the impact of the
DURCA-0, 1 and 2 hypothesis on the EOS properties.

To do so, we calculate the average value of a set of observables hereafter named generically $A$, such as the energy per particle $E/A$, the energy density $\epsilon$, the pressure $P$, 
the symmetry energy $S$, the sound velocity $v_s/c$,  all weighted by the probability $p_{lik}$.
The average value and standard deviation of $A$ are defined as
\begin{eqnarray}
\langle A \rangle &=& \left\{\prod_{\alpha=1}^8 \int dP_\alpha\right\} \; p_{lik} (\{P_{\alpha}\}) A(\{P_{\alpha}\}) \, , \\
\sigma_A^2  &=&  \left\{\prod_{\alpha=1}^8 \int dP_\alpha\right\} \; p_{lik} (\{P_{\alpha}\})  \left[ \langle A \rangle - A(\{P_{\alpha}\}) \right]^2 \, , 
\end{eqnarray}
and are evaluated as a function of the density $n_0$. In the following, we limit the range of densities from $n_{sat}$ up to about $4n_{sat}$. This is the range
of densities which is covered by most of the EOS.

\begin{figure*}[p]
\begin{center}
\includegraphics[angle=0,width=0.7\linewidth]{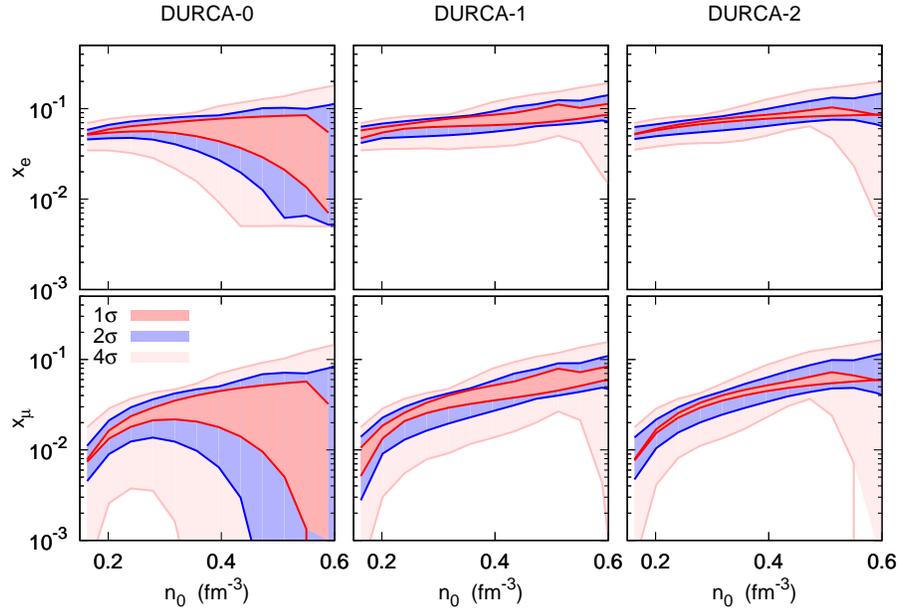}
\end{center}
\caption{(Color online) Same as Fig.~\ref{fig:xp} for the electron fraction $x_e$ and muon fraction $x_{\mu}$.}
\label{fig:xx} 
\end{figure*}

\begin{figure*}[p]
\begin{center}
\includegraphics[angle=0,width=0.7\linewidth]{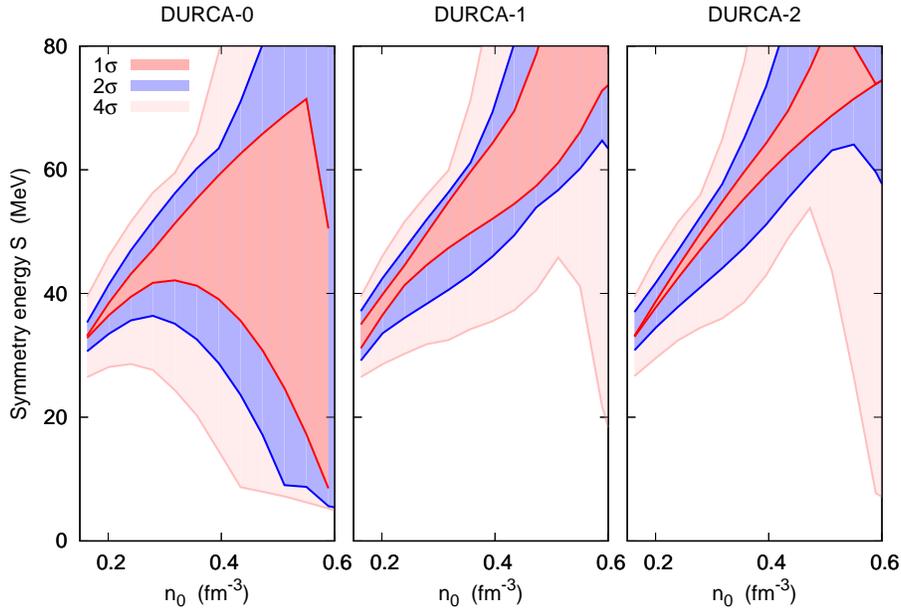}
\end{center}
\caption{(Color online) Same as Fig.~\ref{fig:xp} for the symmetry energy.}
\label{fig:esym} 
\end{figure*}

We show in Fig.~\ref{fig:xp} the density dependence of the distribution of proton fraction for the different scenarii DURCA-0, 1 and 2, as well as
the distribution associated to the threshold condition $x_p^{DURCA}$.
The threshold condition indeed slightly changes with the EOS since it is influenced by the symmetry energy $S(n_0)$.
Fig.~\ref{fig:xp} shows that the threshold condition $x_p^{DURCA}$ has a very narrow distribution and is almost identical for DURCA-1 and 2.
It is however a bit more spread for DURCA-0 hypothesis.
However, the density dependence of  $x_p^{DURCA}$ is rather weak and it is almost independent of the density for $n_0>0.3$~fm$^{-3}$.

The density dependence of the proton fraction $x_p$ is also interesting to analyze.
We can see that the threshold $x_p$ value can be easily overcome for all DURCA hypothesis, at least at the $2\sigma$ level. 
This means that in the corresponding EOS the density domain where $x_p>x_p^{DURCA}$ is never met if the NS mass is below the limiting mass we have supposed for dUrca.
The densities  at which $x_p\approx x_p^{DURCA}$ in each panel correspond approximately to the average central densities for $2M_\odot$, $1.8M_\odot$, and $1.6M_\odot$ NS represented
in Fig.~\ref{fig:impactMR5}.
More quantitatively, for the DURCA-0 hypothesis, the proton fraction remains below the threshold for most of the EOS.
For DURCA-1, the proton fraction reaches the threshold for densities above 0.45~fm$^{-3}$.
And finally, for DURCA-2, the proton fraction reach the threshold for densities above 0.35~fm$^{-3}$.

\begin{figure*}[tb]
\begin{center}
\includegraphics[angle=0,width=0.7\linewidth]{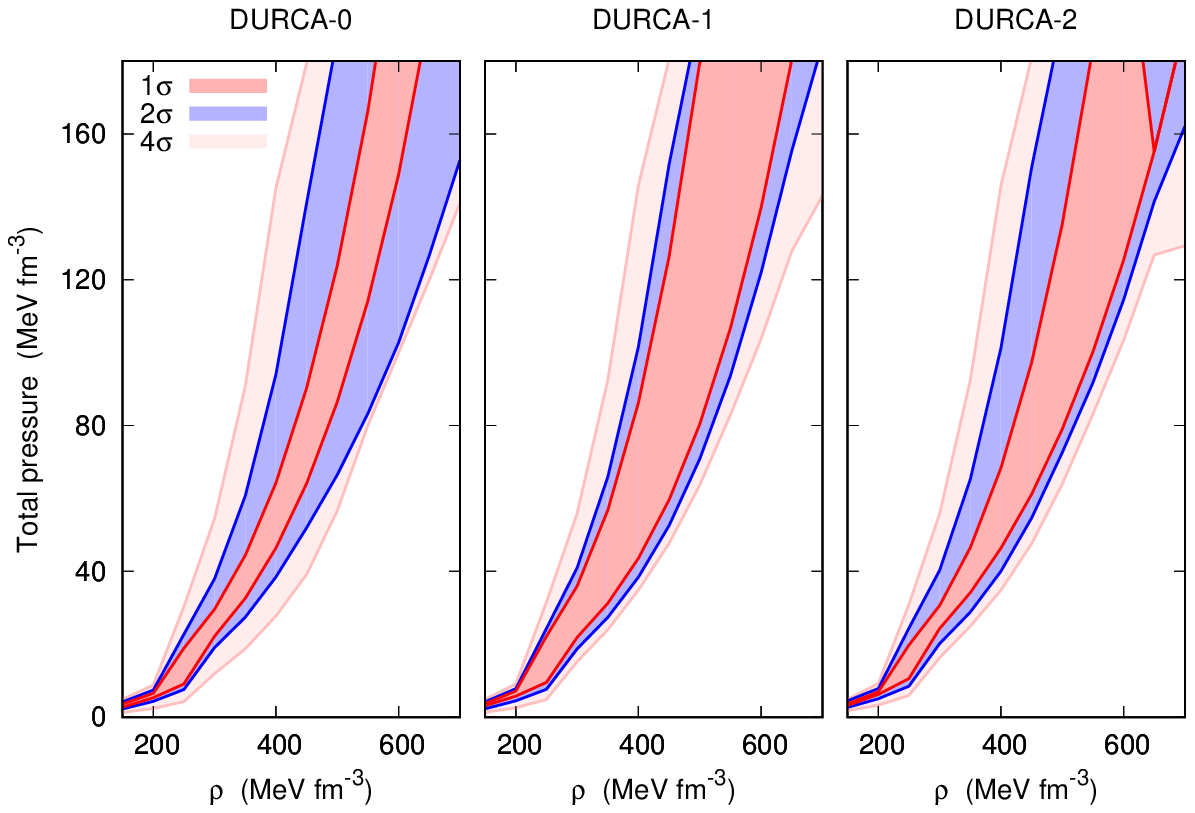}
\end{center}
\caption{(Color online) Total pressure function of the energy density $\rho$ (including the rest mass contribution).}
\label{fig:ptot} 
\end{figure*}

We represent in Fig.~\ref{fig:xx} the probability distributions for the electron fraction $x_e$ (top panels) and the muon fraction $x_\mu$ (bottom panels) as a
function of the density $n_0$, and for the three dUrca hypothesis.
We recall the charge neutrality imposes $x_p=x_e+x_\mu$, and as a matter of fact, the relation still holds approximately for the average, compare Figs.~\ref{fig:xp} and \ref{fig:xx}.
The electron and muon fraction are distributed in a narrow band for DURCA-1 and 2, while they are more widely spread for DURCA-0.
For DURCA-0, very low values for $x_e$ and $x_\mu$ at high density are possible, at variance with DURCA-1 and 2.

%\begin{figure*}[tb]
%\begin{center}
%\includegraphics[angle=0,width=0.7\linewidth]{plot-eos-1-DURCA-E2AB.eps}
%\end{center}
%\caption{(Color online) Same as Fig.~\ref{fig:xp} for the baryon energy per particles at $\beta$-equilibrium.}
%\label{fig:e2ab} 
%\end{figure*}

The density dependence of the symmetry energy as a function of $n_0$ is shown in Fig.~\ref{fig:esym} for the three hypothesis.
It is an interesting quantity since it can be shown that the symmetry energy $S(n_0)$ has a direct impact on the electron fraction~\cite{Lattimer2001a,Steiner2005a}.
As expected, the symmetry energy is softer for DURCA-0 compared to DURCA-1 and 2: For DURCA-1 and 2 the symmetry energy is an increasing function of 
the density $n_0$ while for DURCA-0 the symmetry energy is less stiff and in some cases bends down towards zero at high density.
Let us remind that we excluded EOS with negative symmetry energies.
The density dependence of the symmetry energy is clearly influenced by the dUrca hypothesis, even if a very large spread at high density is still observed, especially for the DURCA-0 hypothesis. 
In particular, it is clear that if we could observationally conclude that dUrca happens in some high mass NS, this would very effectively exclude soft and super-soft behavior for the symmetry energy.

%The probability distribution of the baryonic energy per nucleon at $\beta$-equilibrium is shown in Fig.~\ref{fig:e2ab} as a function of the density $n_0$ and for the three hypothesis
%DURCA-0, 1, and 2.
%The impact of the dUrca hypothesis is very weak, despite the fact that the proton fractions are different.

%\begin{figure*}[tb]
%\begin{center}
%\includegraphics[angle=0,width=0.7\linewidth]{plot-eos-1-DURCA-Pb.eps}
%\end{center}
%\caption{(Color online) Same as Fig.~\ref{fig:xp} for the baryonic pressure.}
%\label{fig:pb} 
%\end{figure*}

%We represent the density dependence of the nucleon pressure at $\beta$-equilibrium in Fig.~\ref{fig:pb}. 
%Here also the three hypothesis concerning the proton fraction have a very limited impact on the pressure.

The EOS, i.e. the total pressure as a function of the total energy density $\rho$ including the rest-mass term, is shown in Fig.~\ref{fig:ptot}.
This quantity, including the contribution of the nucleons and of the leptons (electrons and muons), is used in the TOV equations,  to determine the mass and radius of NS showed above.
Here also, the impact of the dUrca hypothesis is found to be very weak, despite the fact that the proton fractions are different.
% It can be noticed some small influence of the dUrca hypothesis: the meta-EOS predicted by DURCA-0 can be slightly softer than the ones predicted by
% DURCA-1 and 2 hypothesis.
%This effect explains why the radii can be smaller for DURCA-0 hypothesis compared to DURCA-1 and 2, see Fig.~\ref{fig:impactMR1}.
%The observed  weak influence of the dUrca constraint on both the energy and the pressure 
It explains why the global properties of the NS shown in Figs.~\ref{fig:impactMR1} and \ref{fig:impactMR2}
are not very much impacted by these hypothesis, and reflects the universality of the EOS under the charge neutrality and $\beta$-equilibrium conditions~\cite{Blaschke2016}.

%Let us remark that our finding confirms a recent analyzis showing that, for some chosen EOS, the EOS is independent of the
%symmetry energy, which governs the proton fraction~\cite{Blaschke2016}.
 
 \begin{figure*}[p]
\begin{center}
\includegraphics[angle=0,width=0.7\linewidth]{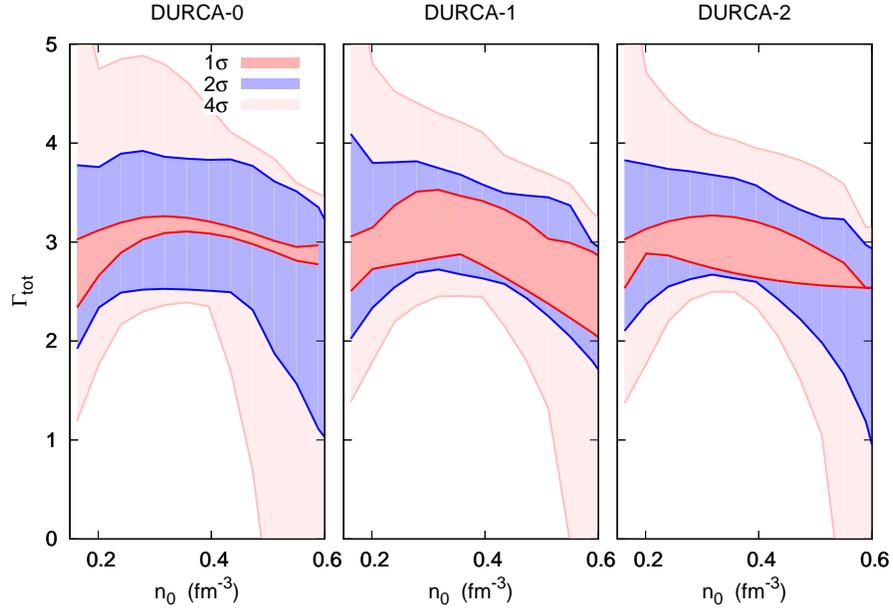}
\end{center}
\caption{(Color online) Adiabatic index $\Gamma(\rho)$ function of the energy density $\rho$ (including the rest mass contribution).}
\label{fig:gamr} 
\end{figure*}

As a complement to the total pressure shown in Fig.~\ref{fig:ptot}, it is interesting to analyze its slope $\Gamma(\rho)$, defined as
 \begin{eqnarray}
 \Gamma(\rho) = \left. \frac{ d \ln P_{tot}}{d \ln \rho} \right\rvert_s \, .
 \end{eqnarray}
$\Gamma(\rho)$ is shown in Fig.~\ref{fig:gamr} as function of the density $n_0$
%total energy density $\rho$ 
and for the three dUrca hypothesis.
%The range of parameters $\Gamma$ are more widely spread for DURCA-0 than for DURCA-1 and 2.
%The upper values at $1\sigma-CL$ of all the dUrca scenarii reasonably match, however, the lower values are smaller for DURCA-0.
%The lower values for DURCA-1 are also slightly lower for DURCA-1 compared to DURCA-2. 
%Notice that the lower the value of $\Gamma$ the softer the EOS.
%Based on the average values obtained for $\Gamma$ at $1\sigma$, we can therefore conclude that most of the EOS satisfying the 
%DURCA-0 condition are softer than the ones satisfying DURCA-1 and 2, and the EOS satisfying DURCA-2 condition seem to be
%on the average the stiffest ones. This conclusion agrees with the behavior of the symmetry energy discussed above.
The density dependence of $\Gamma$ is also rather universal (independent of the dUrca hypothesis).
The average value of $\Gamma$ is between 2 and 4 and it depends weakly of the density.

\begin{figure*}[p]
\begin{center}
\includegraphics[angle=0,width=0.7\linewidth]{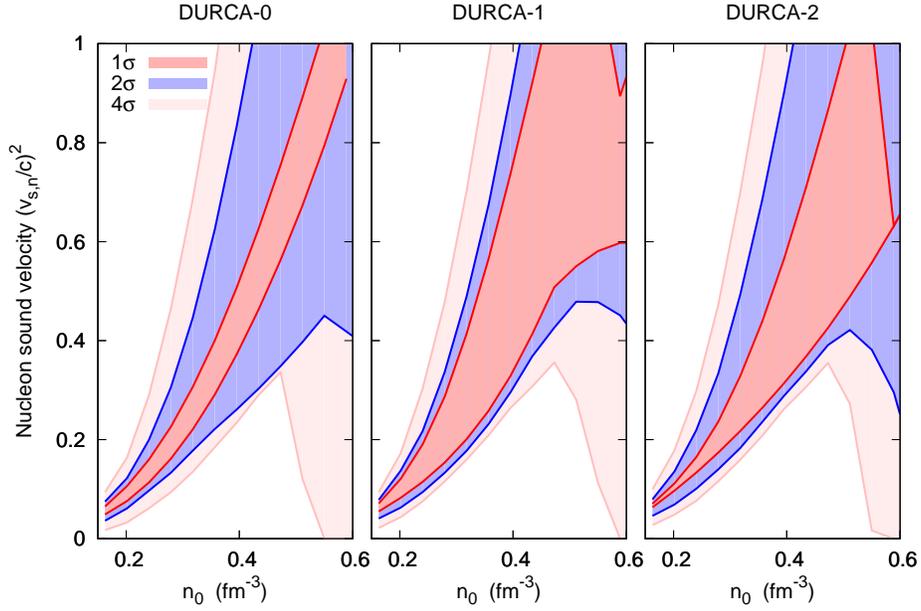}
\end{center}
\caption{(Color online) Same as Fig.~\ref{fig:xp} for the baryon sound velocity.}
\label{fig:vsb} 
\end{figure*}

The last quantity that we analyze is the nucleon sound velocity.
It is the sound velocity calculated from the nucleon pressure and energy per particle as
\begin{eqnarray}
\left(v_{s,n}/c\right)^2 = \frac{d P_{nuc}}{d n_0}/(mc^2+e_{nuc}+P_{nuc}/n_0)
\label{eq:vsn}
\end{eqnarray}
Notice the difference between Eq.~(\ref{eq:vsn}) for nucleons only and Eq.~(\ref{eq:vs}) for the total system including nucleons
and leptons.
The contribution of the leptons increase the sound velocity by about 10-15\%.
It is interesting to represent the quantity $\left(v_{s,n}/c\right)^2$ since it can be compared to the sound velocity usually associated to
nucleonic EOS for symmetric and asymmetric matter, but not necessarily at $\beta$-equilibrium, see for instance Refs.~\cite{Su1988,Bedaque2014a}.
The nucleon sound velocity $\left(v_{s,n}/c\right)^2$ is represented in Fig.~\ref{fig:vsb} as a function of the density $n_0$ and for the three dUrca
hypothesis.
The density dependence of the sound velocity has recently been discussed with respect to its expected limit $1/3$ at very high density, when
matter is composed of a free gas of non-interacting quarks.
It can be shown in perturbation theory that this limit is reached from below as the density increases~\cite{Bedaque2014a}.
Since the sound velocity starts with a positive slope around saturation density, and becomes larger than $1/3$ on average for densities below 2-3$n_{sat}$,
the asymptotic limit implies that the sound velocity has to bend down at least one time.
At high density and for soft EOS the slope of the sound velocity can bend down in nuclear matter.
So the bending down of the sound velocity at high density does not necessarily require specific features, such as phase transition to quark matter, and can be also be obtained for a simple nucleonic EOS.

In summary of this section, we have analyzed some features of the meta-EOS and their link with the dUrca hypothesis.
While the dUrca hypothesis influences strongly the particle fractions, reflecting different density dependence of the symmetry energy,
the EOS is almost independent of the dUrca hypothesis.
Our analysis confirms the universal behavior of the EOS discussed in Ref.~\cite{Blaschke2016} and  generalizes it to cases where dUrca is allowed for large masses.
%DURCA-0 allows softer EOS compared to DURCA-1 and 2.
%For this reason, the radius of NS can be slightly smaller with DURCA-0 compared to DURCA-1 and 2.
%The DURCA-0 EOS is also asy-softer than DURCA-1 and 2.
%This feature is reflected on the particle fraction: the electron fraction is smaller for DURCA-0.

%%%%%%%%%%%%%%%%%%%%%%%%%%%%%%%%%%%%%%%%%%%%%%%%%%%%%
\section{Inversion problem: what is the best EOS reproducing a given mass-radius relation?}
\label{sec:inversion}
%%%%%%%%%%%%%%%%%%%%%%%%%%%%%%%%%%%%%%%%%%%%%%%%%%%%%

In the previous sections, we have deduced the mass-radius (MR) relation from a set of meta-EOS.
In this section, we illustrate the use of  Bayesian analysis to solve the inversion problem: given a MR relation,
how to extract the best meta-EOS passing through?

First, we need a set of data which is the MR relation to fit.
Let $R_{data}(M_{k_M})$ be the set of MR relations,  $\sigma_{data}(M_{k_M})$ the associated error-bar in the
radius  , and $N_M$ the number of data to fit. Then we can define the $\chi_{MR,i}^2$ 
function as:
\begin{eqnarray}
\chi_{MR,i}^2 = \frac 1 {N_M} \sum_{k_M=1}^{N_M} \left( \frac{ R_{i}(M_{k_M}) - R_{data}(M_{k_M}) }{\sigma_{data}(M_{k_M})}\right)^2 \, .
\label{eq:chir}
\end{eqnarray}
This quantity evaluates the goodness of a given meta-EOS (represented by its associated set of parameters $i=\{P_\alpha\}$), where $R_i(M_{k_M})$ 
is the MR relation of the EOS $i$. 

The associated likelihood probability is
\begin{equation}
p_{lik,MR}(i)=N_{lik,MR}^{-1}\, \exp \left( -\frac 1 2 \chi_{MR,i}^2\right) \, .
\label{eq:pmr}
\end{equation}

Solving the inversion problem consists in analyzing the distribution of the likelihood probability
$p_{lik,MR}(i)$ for each meta-EOS $i$ and extract the more probable parameters, and their uncertainties.
A 1-parameter probability $p_{1,MR}$ can be deduced from the multi-parameter probability $p_{lik,MR}$, as
\begin{eqnarray}
p_{1,MR}(P_\alpha)  &=& \left\{\prod_{\beta(\ne\alpha)=1}^8 \int dP_\beta\right\} \; p_{lik,MR} (i) \, ,  \\
\end{eqnarray}
and the centroid $\langle P_{\alpha,MR}\rangle$ and standard deviation $\sigma_{\alpha,MR}$ of the 1-parameter probability 
$p_{1,MR}$, are calculated from the probability distribution in the standard way:
\begin{eqnarray}
\langle P_{\alpha,MR}\rangle &=& \int dP_{\alpha} P_\alpha p_{1,MR}(P_\alpha)  \, , \label{eq:pmean}\\
\sigma_{\alpha,MR}^2 &=&   \int dP_{\alpha} \left[ \langle P^R_\alpha\rangle - P_\alpha\right]^2 p_{1,MR}(P_\alpha) 
 \, , \nonumber \\
&=& \langle P_\alpha^2 \rangle- \langle P_{\alpha,R} \rangle^2 \label{eq:pstd}.
\end{eqnarray}
From Eqs.~(\ref{eq:pmean})-(\ref{eq:pstd}) one can deduce the best set of parameters (and their associated dispersion) 
which reproduce the data.

We will illustrate this method in the following subsections: first we will analyze the most probable MR relations obtained in Sec.~\ref{sec:impactns},
and second, we will analyze the impact of shifting the more probable MR to smaller radii on the empirical parameters.

\begin{figure*}[tb]
\begin{center}
\includegraphics[angle=0,width=0.7\linewidth]{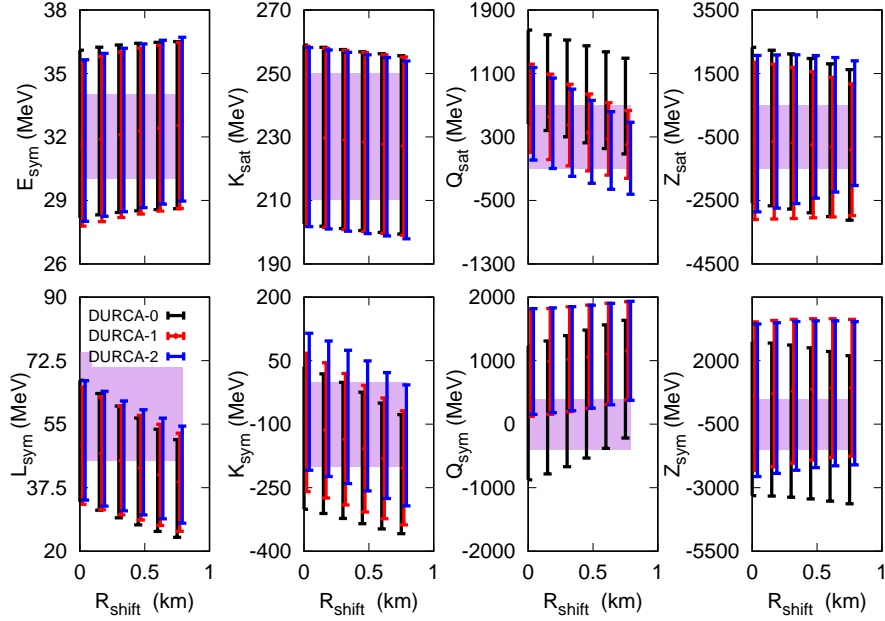}
\end{center}
\caption{(Color online) Effect of shifting the average radius by $-R_{shift}$ on the empirical parameters. 
The purple 1-$\sigma$ bands shown for each empirical parameter are deduced from Tab.~\ref{tab:epbound}.
See text for more details.}
\label{fig:shift} 
\end{figure*}

\subsection{Analysis of the most probable MR relations}
\label{sec:optimizedparam}

In this section, we extract the best meta-EOS which reproduce the average MR relationships obtained in Sec.~\ref{sec:impactns}.
We therefore run over the 25 million meta-EOS generated in Sec.~\ref{sec:impactns} and associate to each of them a new probability
$p_{lik,MR}$ defined from Eq.~(\ref{eq:pmr}) where the data is the radius and its $1\sigma$ width given in Fig.~\ref{fig:impactMR3}.

The results based on $p_{1,MR}$ are shown in Tab.~\ref{tab:epcentroids} and can be compared to the prior distribution and to
the distribution of parameters deduced from the original likelihood probability $p_{1}$ (containing the physical constraints on the
causality, stability and symmetry energy).
There is a good agreement between the low order empirical parameters determined from $p_{1}$ and from $p_{1,MR}$ such as
$E_{sym}$, $L_{sym}$, $K_{sat/sym}$.
For the higher order empirical parameters, such as $Q_{sat/sym}$ and $Z_{sat/sym}$ the centroid are still quite compatible 
between $p_{1}$ and $p_{1,MR}$ (considering the $1\sigma$-CL).
The uncertainties associated to $Q_{sat/sym}$ and $Z_{sat/sym}$ from the probability distribution $p_{1,MR}$ are however large, 
even a bit larger than the original uncertainty defined in the prior.
It is a sign that these parameters are not well constrained by the MR relation at $1\sigma$-CL, since they constrain the high density
domain of the EOS and thus, they weakly impact the MR relation below $2M_{\odot}$.

\subsection{Impact of shifting the more probable MR relation to smaller radii}
\label{sec:shift}

In this section, we address another question of importance: suppose that the radius of neutron stars is once measured and found 
to be smaller than our prediction;
which parameters will be  mostly impacted by such a measurement?

In other word, we want to analyze the correlation between the best parameter set and the average position of the radius. 
To do so, we consider that the radius is uniformly shifted down as, $R_{data}(M) = \langle R(M) \rangle - R_{shift}$, 
where $\langle R(M) \rangle$ is taken from Fig.~\ref{fig:impactMR3} as well as the width $\sigma_{data}=\sigma_R(M)$,
which is not modified in this example.

We should remark that the hypothesis of constant shift with no modification of width is not fully realistic. 
Indeed the high density EOS, explored in the most massive neutron stars, 
is more uncertain than the low density one, meaning that the radii corresponding to the lighter NS are in principle better constrained.
However, the universal behavior observed in Figs.\ref{fig:impactMR1},\ref{fig:impactMR2} suggests that this schematic example 
can still give significant information on the parameters
which are the most influential in a radius determination.  

Fig.~\ref{fig:shift} shows the impact on the empirical parameters, of shifting the average radius down to about 1~km .
As expected from our previous analyzes, the empirical parameters $E_{sym}$, $K_{sat}$, $Z_{sat}$ and $Z_{sym}$ are almost 
insensitive to the shift of the radius.
$E_{sym}$ is weakly impacted because the baryon pressure is independent of it.
$K_{sat}$ has a weak impact because it is sufficiently well known and varies only in a small interval.
$Z_{sat}$ and $Z_{sym}$ have weak impact because they influence the pressure at densities which are higher than the one which matter here.
The more impacted empirical parameters are $L_{sym}$, $K_{sym}$, $Q_{sat}$ and $Q_{sym}$.
As the shift increases (the total radius decreases), the empirical parameters $L_{sym}$ and $K_{sym}$ decreases.
This result is expected since these empirical parameters are the more influential on the pressure around saturation density: the pressure
is proportional to $L_{sym}$, while $K_{sym}$ governs the density dependence of the pressure at the lowest order.
Finally $Q_{sat}$ and $Q_{sym}$ impact the density dependence of the pressure at higher density than $K_{sym}$ (second order).
They are sensitive to the MR relation for high mass NS.
A lower radius for high mass NS requires a softening of the EOS, which implies a decrease of $Q_{sat}$ as expected from our previous
analysis.
The effect of this softening is however partially compensated by $Q_{sym}$ as see in Fig.~\ref{fig:shift}.

In summary, the empirical parameters which are the most impacted by the fit to lower radii are mainly $L_{sym}$, $K_{sym}$, $Q_{sat}$ 
and $Q_{sym}$.

\section{Conclusions}
\label{Sec:Conclusions}

In this paper, we have applied the meta-EOS presented in paper~I~\cite{Margueron2017a} to zero temperature $\beta$-equilibrium neutron stars, assuming they are only composed of nucleons, electrons and muons.
We have first performed a simple sensitivity analysis varying the empirical parameters independently in order to study
their impact on the MR relation.
The empirical parameters $L_{sym}$, $K_{sym}$, $Q_{sat}$ are found to be the more important ones.
A better determination of these empirical parameter will reduce the error-bars on the MR relation predicted by nucleonic EOS.

We have also performed a Bayesian analysis, taking as a prior the estimated empirical parameters average value and uncertainty
determined in paper~I, and filtering among the approximately 25~million generated meta-EOS the ones which satisfy the basic physical
requirements of causality, stability and positiveness of the symmetry energy in a density interval corresponding to NS up
to $2M_\odot$.
We also divided the meta-EOS into three groups according to their prediction for the mass range where dUrca may occur: no dUrca
up to $2M_\odot$ (DURCA-0), dUrca for NS with masses between $1.8M_\odot$ and $2M_\odot$ (DURCA-1), and finally
dUrca for NS with threshold masses between $1.6M_\odot$ and $1.8M_\odot$ (DURCA-2).
We found that the final influence of the physical filtering and dUrca process on the probability distribution of empirical parameters
(the posterior) is rather weak.
The most impacted empirical parameters are $L_{sym}$, $K_{sym}$, $Q_{sat/sym}$ and the centroid of $K_{sym}$ is clearly
increased from DURCA-0 to DURCA-2.
The correlation between the empirical parameters revealed only very weak correlations, suggesting that most correlations observed in the literature originate from the lack of flexibility of existing phenomenological functionals, {or additional constraints that we have not considered here, such as for instance the experimental masses and charge radii of finite nuclei}.

We have also used the probability distribution of parameters to quantitatively predict the confidence intervals on global properties of NS, such as their radius, momentum
of inertia, surface redshift, central proton fraction, crust thickness and central density as function of the mass and of the dUrca
hypothesis.
%While these quantities are quite insensitive to the dUrca hypothesis on the average, we noticed that NS with small radius, slightly lower than 11~km 
%for instance, could only be explained by the DURCA-0 hypothesis, but very marginally (at more than $4\sigma$-CL from the most probable
%distribution of radius).
The central proton fraction is substantially impacted by the dUrca hypothesis, as expected, and the central density can be
larger for DURCA-0 hypothesis compared to DURCA-1 and 2.
The EOS is however found to exhibit an universal behavior against the dUrca hypothesis under the condition of charge neutrality and $\beta$-equilibrium.
If composed exclusively of nucleons and leptons, our prediction is that neutron stars have a radius of 12.7$\pm$0.4~km for masses between 1 and $2M_{\odot}$.

Assuming low compactness NS are only composed of nucleons and leptons we could use our predictions at $1\sigma$-CL to correlate a measurement of compactness (for instance 0.105$\pm$0.002~$M_\odot$~km$^{-1}$ proposed for RX J0720.4-3125~\cite{Hambaryan2017}) to a prediction of its radius (12.7$\pm0.3$~km) and mass (1.33$\pm$0.04~$M_\odot$).
These predictions are done without any assumption on the functional form of the EOS, and with the only requirement that the EOS is nucleonic and satisfies basic physical constraints. 
As such, the prediction can be qualified as model independent.

We have discussed in great details the meta-EOS at $\beta$-equilibrium, as predicted by the posterior probability distribution and we discussed the differences induced by the dUrca scenario.
The proton, electron and muon fractions are clearly impacted by the dUrca scenario.
This can be related to the density dependence of the symmetry energy.
DURCA-0 hypothesis produces a more asy-soft EOS than DURCA-1 and 2.
The EOS, $P(\rho)$ as well as its logarithmic derivative $\Gamma(\rho)$, confirm the 
%fact that DURCA-0 is globally softer than DURCA-1 and 2.
universal behavior predicted in Ref.~\cite{Blaschke2016} for EOS without dUrca. 
We extend this prediction for EOS where dUrca is allowed for high mass NSs. 
Finally, we represented the probability distribution of nucleon sound velocity and discussed its expected asymptotic limit.

The last part of this work addresses the question of the inverse problem: how does an improved knowledge of the EOS can be obtained from accurate measurements of NS masses and radii.
We have shown that the empirical parameters $L_{sym}$, $K_{sym}$, and $Q_{sat/sym}$ are the most impacted by the measurement of the NS radii.

In conclusion, the empirical parameters encode very important properties of nuclear matter from which accurate predictions can be performed. 
They can include up-to-date constraints from experimental data as well as ab-initio approaches, 
and probe the accuracy of the predictions for dense matter EOS based on the present knowledge.
In the present work, we have pointed out the most important empirical parameters which will require more attention in the future: $L_{sym}$, $K_{sym}$, and $Q_{sat/sym}$.
They are mainly responsible to the uncertainty in the MR relation based on nucleonic EOS.
Finite nuclei may also provide better constraints on some empirical parameters, such as the lowest order ones.
In the future, we plan to apply the meta-EOS to the description of the global properties of finite nuclei, such as their masses and radii, from the density functional approach.
We also plan to continue on our analysis by including some additional constraints on the density dependence of the energy per particle and of the symmetry energy. 
These additional constraints can easily be included in our selection filter of the likelihood probability.

\section{Acknowledgments}

This work was partially supported by CNPq (Brazil), processes (209440/2013-9) and (400877/2015-5), by the SN2NS project ANR-10-BLAN-0503 and by New-CompStar COST action MP1304.
One of us (J.M.) wants to thank Nicolas Baillot, S\'ebastien Guillot, Jim Lattimer, Sanjay Reddy and Ingo Tews for stimulating discussions and interesting suggestions during the completion of this work.
This work also benefited from the INT programs "Bayesian analysis for nuclear physics" INT-16-2a and "Phase of Dense Matter" INT-16-2b.
The numerical part of this work has solicited the super-computer facility of CC-IN2P3 (about 500 CPU during 1 month).

\bibliographystyle{apsrev}

\end{document}